\documentclass[a4paper,11pt]{article}
\usepackage{amsfonts}
\usepackage{amsmath}
\usepackage[a4paper]{hyperref}
\usepackage{graphicx}

\input{tcilatex}
\def\Qlb#1{#1}

\def\Qcb#1{#1} 

\def\FRAME#1#2#3#4#5#6#7#8
{
 \begin{figure}[ptbh]
 \begin{center}
 \includegraphics[height=#3]{#7}
 \caption{#5}
 \label{#6}
 \end{center}
 \end{figure}
}

\begin{document}

\title{Minisuperspace Models in M-theory}
\author{Sergey Grigorian \\
%EndAName
\\
DAMTP\\
Centre for Mathematical Sciences\\
Wilberforce Road\\
Cambridge CB3 0WA\\
United Kingdom}
\date{22 July 2007}
\maketitle

\begin{abstract}
We derive the full canonical formulation of the bosonic sector of
11-dimensional supergravity, and explicitly present the constraint algebra.
We then compactify M-theory on a warped product of homogeneous spaces of
constant curvature, and construct a minisuperspace of scale factors. First
classical behaviour of the minisuperspace system is analysed, and then a
quantum theory is constructed. It turns out that there similarities with the
\textquotedblleft pre-Big Bang\textquotedblright\ scenario in String Theory.
\end{abstract}

\section{Introduction}

One of the most fundamental problems in theoretical physics is the search
for a quantum theory which would unify gravity with other interactions. Over
the past 20 years, superstring theory emerged as a successful candidate for
this role. It was later discovered that all five superstring theories can
all be obtained as special limits of a more general eleven dimensional
theory known as M-theory and moreover, the low energy limit of which is the
eleven dimensional supergravity \cite{Witten:1995ex,Townsend:1995kk}. The
complete formulation of M-theory is however not known yet.

In a cosmological context, there is another approach to quantum gravity
which was pioneered by DeWitt in \cite{BSDeWitt1}. Here, the gravitational
action is reformulated as a constrained Hamiltonian system and then
quantized canonically. The resulting wavefunction is sometimes referred to
as the \textquotedblleft wavefunction of the universe\textquotedblright\ 
\cite{HartleHawking}, as it describes the state of the universe. Such a
wavefunction is a function on the superspace - an infinite dimensional space
of all possible metrics modulo the diffeomorphisms. Although this procedure
of course does not give a full theory of quantum gravity, it does give a low
energy approximation, which is enough to capture some quantum effects such
as tunnelling \cite{HartleHawking,Vilenkin:1987kf}. Since the behaviour of
the wavefunction in the full infinite-dimensional superspace is difficult to
analyze, models with a reduced number of degrees of freedom have been
considered. In these models only a finite subset of the original degrees of
freedom are allowed to vary, while the rest are fixed, so that the
wavefunction becomes a function on a finite-dimensional \emph{minisuperspace}%
. In the early Universe, it is perceived that quantum gravity effects should
become important, and so such minisuperspace models, where the degrees of
freedom are the spatial scale factor in the Friedmann-Robertson-Walker (FRW)
metric and possibly a scalar field, have been used to explore different
quantum cosmological scenarios \cite{HartleHawking,Vilenkin:1987kf,
Hawking:1983hj, Rubakov:1984bh,Vilenkin:1986cy}.

With the advent of superstring theory, the above ideas have been applied in
the context of superstring theory \cite{BentoBertolami,
GasperiniVeneziano96a, GasperiniVeneziano96b, Dabrowski}\cite%
{Goldwirth:1993ha},\cite{Cavaglia:1999ka}. This time however, the starting
point is the lowest order string effective action, possibly with a dilaton
potential or a cosmological constant put in. Hence compared to the pure
gravity case, there are new degrees of freedom - the dilaton and any of the
tensor fields that appear. The minisuperspace models studied in the quantum
string cosmology setting now have the FRW scale factor and the dilaton field
as the independent degrees of freedom. In particular, progress has been made
in quantum string cosmological description of the \textquotedblleft pre-Big
Bang scenario\textquotedblright\ \cite{GasperiniVeneziano96a,
GasperiniVeneziano96b, Dabrowski,GasperiniVeneziano02}. In this scenario,
the universe evolves from a weakly coupled string vacuum state to a FRW
geometry through a region of large curvature. Classically there is the
problem that the pre-Big Bang and post-Big Bang branches are separated by a
high-curvature singularity. However, in the minisuperspace model for a
spatially flat ($k=0$) FRW space-time with a suitable dilaton potential, it
is possible to find a wavefunction which allows tunnelling between the two
classically disconnected branches and this solves the problem of transition
between the two regimes.

With M-theory being a good candidate to be a \textquotedblleft theory of
everything\textquotedblright , it is interesting to see what the canonical
quantization of the low energy effective theory can give, given the
achievements of this approach in the pure gravity and superstring theory
contexts. In Section 2 below, we start with the bosonic action for
eleven-dimensional supergravity and reformulate the theory as a canonical
constrained Hamiltonian system. The canonical formulation of
eleven-dimensional supergravity has been considered before in \cite%
{Diaz:1986jw} and \cite{Diaz:1986jx}, but here we explicitly give the
constraint algebra, at least for the bosonic constraints.

Then in Section 3, we reduce the system to a minisuperspace model. This is
done by restricting the metric ansatz so that its spatial part is a warped
product of a number of homogeneous spaces of constant curvature, and the
supergravity $4$-form is also restricted so that only its $4$-space
components are allowed to be non-zero. In the case when only one of the
spatial components has non-vanishing curvature and the $4$-form vanishes
completely, it is possible to solve exactly both the classical equations of
motion and the corresponding equation for the wavefunction.

In section 4, we consider the classical and quantum solutions in the cases
of vanishing, negative and positive spatial curvature. Spatially flat
M-theory minisuperspace models have also been considered in \cite%
{CavagliaMoniz} and \cite{Cavaglia:2001sb}. We carefully consider these
systems, fixing the gauge and paying attention to self-adjointness of the
Hamiltonian. It turns out that the positive and negative curvature cases can
exhibit very similar behaviour to the string theory minisuperspace models
described above with negative and positive dilaton potentials in the
Hamiltonian respectively. However it also turns out that in the negative
curvature case, the boundary condition which leads to a tunnelling effect,
in fact also leads to lost self-adjointness of the Hamiltonian. This is a
purely mathematical consequence, and certainly deserves further
investigation to find out the what is the correct physical reason behind
this.

In section 5, we look at the case where the $4$-form is switched on, the $3$%
-space is flat, and one other spatial component is of positive curvature.

We will be using the following conventions. The spacetime signature will be
taken as $\left( -++...+\right) $ and all the curvature conventions are the
same as in \cite{MTW}. Greek indices $\mu ,\nu ,\rho ,...$ range from $0$ to 
$10,$ while the indices $\alpha ,\beta ,\gamma ,...$ range from $0$ to $3$.
Latin indices $a,b,c,...$ range from $1$ to $10$. The units used are such
that $\hbar =c=16\pi G^{\left( 11\right) }=1$.

\section{Canonical formulation}

In this section we set up the canonical formalism for the bosonic sector of $%
11$-dimensional supergravity, with the field content being just the metric $%
\hat{g}_{\mu \nu }$ and the $3$-form potential $\hat{A}$, with field
strength $F=d\hat{A}$.

The action for the bosonic fields is \cite{Cremmer:1978km}: 
\begin{eqnarray}
S &=&\int d^{11}x\left( -\hat{g}\right) ^{\frac{1}{2}}R^{\left( 11\right) }-%
\frac{1}{2}\int F\wedge \ast F-\frac{1}{6}\int \hat{A}\wedge F\wedge F 
\notag \\
&=&\int d^{11}x\left( -\hat{g}\right) ^{\frac{1}{2}}\left( R^{\left(
11\right) }-\frac{1}{48}F^{\mu _{1}...\mu _{4}}F_{\mu _{1}...\mu _{4}}-\frac{%
1}{12^{4}}\varepsilon ^{\mu _{1}...\mu _{11}}\hat{A}_{\mu _{1}\mu _{2}\mu
_{3}}F_{\mu _{4}...\mu _{7}}F_{\mu _{8}...\mu _{11}}\right)  \label{sugra2}
\end{eqnarray}%
where $\hat{g}=\det \left[ \hat{g}_{\mu \nu }\right] $ and $F_{\mu
_{1}...\mu _{4}}=4\partial _{\lbrack \mu _{1}}\hat{A}_{\mu _{2}\mu _{3}\mu
_{4}]}$. The $11$-dimensional alternating tensor $\varepsilon ^{\mu
_{1}...\mu _{11}}$ is defined by 
\begin{eqnarray*}
\varepsilon ^{\mu _{1}...\mu _{11}} &=&\left( -\hat{g}\right) ^{-\frac{1}{2}%
}\eta ^{\mu _{1}...\mu _{11}} \\
\varepsilon _{\mu _{1}...\mu _{11}} &=&\left( -\hat{g}\right) ^{\frac{1}{2}%
}\eta _{\mu _{1}...\mu _{11}}
\end{eqnarray*}%
where $\eta ^{\mu _{1}....\mu _{11}}=-\eta _{\mu _{1}...\mu _{11}}$ is the
alternating symbol.

To decompose the metric into spatial and temporal parts, we use the
following ansatz \cite{BSDeWitt1}:%
\begin{equation}
\hat{g}_{\mu \nu }=\left( 
\begin{array}{cc}
-\alpha ^{2}+\beta _{a}\beta ^{a} & \beta _{a} \\ 
\beta _{b} & \gamma _{ab}%
\end{array}%
\right) .  \label{metricansatz}
\end{equation}%
The inverse metric is given by 
\begin{equation}
\hat{g}^{\mu \nu }=\left( 
\begin{array}{cc}
-\alpha ^{-2} & \alpha ^{-2}\beta ^{a} \\ 
\alpha ^{-2}\beta ^{b} & \gamma ^{ab}-\alpha ^{-2}\beta ^{a}\beta ^{b}%
\end{array}%
\right)  \label{invmetric}
\end{equation}%
where $\gamma _{ac}\gamma ^{bc}=\delta _{a}^{b}$, and $\beta ^{a}=\gamma
^{ab}\beta _{b}$.

Using this ansatz, we follow \cite{MTW} to express canonically the
gravitational action.

Consider a hypersurface $\Sigma ,$ given by $t=const$. The future-pointing
normal vector $n^{\mu }$ to this hypersurface is given by 
\begin{equation}
n^{\mu }=\left( 
\begin{array}{c}
\alpha ^{-1} \\ 
-\alpha ^{-1}\beta ^{a}%
\end{array}%
\right)  \label{normal}
\end{equation}%
and the corresponding covector is $n_{\mu }=\left( -\alpha ,\mathbf{0}%
\right) $, so hence $n_{\mu }n^{\mu }=-1$.

The second fundamental form $K_{\mu \nu }$ for $\Sigma $ is defined by

\begin{equation}
K_{\mu \nu }=-h_{\mu }^{\rho }h_{\nu }^{\sigma }n_{\rho ;\sigma }
\label{2ndfunforma}
\end{equation}%
where the semicolon denotes covariant differentiation with respect to the
11-dimensional metric $\hat{g}$ and $h_{\mu }^{\rho }$ is the projector onto 
$\Sigma $ defined by 
\begin{equation*}
h_{\mu \nu }=g_{\mu \nu }+n_{\mu }n_{\nu }.
\end{equation*}%
Note that the sign in (\ref{2ndfunforma}) depends on the convention used, so
here we follow \cite{MTW}.

From (\ref{2ndfunforma}) we have in particular 
\begin{equation}
K_{ab}=-n_{a;b}.  \label{2ndfunform}
\end{equation}%
Using the definition of $n^{\mu }$ (\ref{normal}), it can be shown that \cite%
{MTW} 
\begin{equation}
K_{ab}=\frac{1}{2}\alpha ^{-1}\left( \beta _{a|b}+\beta _{b|a}-\gamma
_{ab,0}\right)  \label{2ndfunform2}
\end{equation}%
where $|$ denotes covariant differentiation with respect to the metric $%
\gamma _{ab}$. Using the Gauss-Codazzi equation (\ref{codazzi}) below, the
full eleven-dimensional curvature can be expressed in terms the intrinsic
curvature of the hypersurface (that is, the curvature of the metric $\gamma
_{ab}$) and the second fundamental form: 
\begin{equation}
R^{\left( 11\right) }=R^{\left( 10\right) }-K_{ab}K^{ab}+K^{2}-2n^{\mu
}n^{\nu }R_{\mu \nu }^{\left( 11\right) }  \label{codazzi}
\end{equation}%
where $K=\gamma ^{ab}K_{ab}$ and $K^{ab}=\gamma ^{ac}\gamma ^{bd}K_{cd}.$
Hence the gravitational Lagrangian density $\mathcal{L}_{grav}$ is given by 
\begin{eqnarray}
\mathcal{L}_{grav} &=&\left( -\hat{g}\right) ^{\frac{1}{2}}R^{\left(
11\right) }  \notag \\
&=&\alpha \gamma ^{\frac{1}{2}}\left( R^{\left( 10\right)
}-K_{ab}K^{ab}+K^{2}-2n^{\mu }n^{\nu }R_{\mu \nu }^{\left( 11\right) }\right)
\notag \\
&=&\alpha \gamma ^{\frac{1}{2}}\left( R^{\left( 10\right)
}+K_{ab}K^{ab}-K^{2}\right) +\text{total derivative terms}  \label{lgrav}
\end{eqnarray}%
where $\gamma =\det \left( \gamma _{ab}\right) $. In the action, the full
derivative terms give rise to a surface integral. We neglect it, since it
does not affect the dynamics of the system.

We now decompose the $3$-form $\hat{A}_{\mu \nu \rho }$ as%
\begin{eqnarray}
\hat{A}_{0ab} &=&B_{ab}  \label{aoij} \\
\hat{A}_{abc} &=&A_{abc}  \label{aijk}
\end{eqnarray}%
and correspondingly, 
\begin{eqnarray}
F_{abcd} &=&4\partial _{\lbrack a}A_{bcd]}  \label{fijkl} \\
F_{0abc} &=&\partial _{0}A_{abc}-3\partial _{\lbrack a}B_{bc]}.
\label{f0ijk}
\end{eqnarray}%
The $F^{2}$ term from the action (\ref{sugra2}) is decomposed as 
\begin{equation}
F^{\mu _{1}...\mu _{4}}F_{\mu _{1}...\mu _{4}}=F_{abcd}F^{abcd}-4F_{\perp
bcd}F_{\perp }^{\ \ bcd}  \label{fsqexp3}
\end{equation}%
where%
\begin{equation}
F_{\perp bcd}=n^{\mu }F_{\mu bcd}=\alpha ^{-1}\left( F_{0bcd}-\beta
^{a}F_{abcd}\right) .  \label{fperp1}
\end{equation}%
Looking at the Chern-Simons term $A\wedge F\wedge F$, we have 
\begin{align}
\eta ^{\mu _{1}...\mu _{11}}\hat{A}_{\mu _{1}\mu _{2}\mu _{3}}F_{\mu
_{4}...\mu _{7}}F_{\mu _{8}...\mu _{11}}& =\eta ^{a_{1}...a_{10}}\left(
12^{2}B_{a_{1}a_{2}}\partial _{\lbrack a_{3}}A_{a_{4}a_{5}a_{6]}}\partial
_{\lbrack a_{7}}A_{a_{8}a_{9}a_{10]}}\right.  \label{lform} \\
& \left. +8\eta ^{a_{1}...a_{10}}\left( \partial
_{0}A_{a_{1}a_{2}a_{3}}\right)
A_{a_{4}a_{5}a_{6}}F_{a_{7}a_{8}a_{9}a_{10}}\right) +\text{total derivative
term}  \notag
\end{align}%
Again, we neglect the total derivative term, since it does not affect the
equations of motion.

Bringing together (\ref{lgrav}), (\ref{fsqexp3}) and (\ref{lform}), we thus
have the total Lagrangian 
\begin{equation*}
L_{tot}=\int d^{10}x\left( \mathcal{L}_{grav}+\mathcal{L}_{form}\right)
\end{equation*}%
where 
\begin{eqnarray}
\mathcal{L}_{grav} &=&\gamma ^{\frac{1}{2}}\alpha \left( R^{\left( 10\right)
}+K_{ab}K^{ab}-K^{2}\right)  \label{lgrav2} \\
\mathcal{L}_{form} &=&\gamma ^{\frac{1}{2}}\left[ -\frac{\alpha }{48}%
F_{abcd}F^{abcd}+\frac{\alpha }{12}F_{\perp bcd}F_{\perp }^{\ \ bcd}\right.
\label{lform2} \\
&&\left. -\varepsilon ^{a_{1}...a_{10}}\left( \frac{1}{12^{2}}%
B_{a_{1}a_{2}}\partial _{\lbrack a_{3}}A_{a_{4}a_{5}a_{6]}}\partial
_{\lbrack a_{7}}A_{a_{8}a_{9}a_{10]}}-\frac{8}{12^{4}}\left( \partial
_{0}A_{a_{1}a_{2}a_{3}}\right)
A_{a_{4}a_{5}a_{6}}F_{a_{7}a_{8}a_{9}a_{10}}\right) \right]  \notag
\end{eqnarray}

We see that the canonical fields in this system are $\alpha ,\beta
^{a},\gamma ^{ab}$ which come from the gravitational Lagrangian, together
with $A_{abc}$ and $B_{ab}$ which come from $\mathcal{L}_{form}$. From the
Lagrangian densities (\ref{lgrav2}) and (\ref{lform2}) we can now write down
the canonical momenta conjugate to these variables: 
\begin{subequations}
\begin{eqnarray}
\pi &=&\frac{\partial \mathcal{L}_{tot}}{\partial \alpha _{,0}}=0
\label{mom1} \\
\pi ^{a} &=&\frac{\partial \mathcal{L}_{tot}}{\partial \beta _{a,0}}=0
\label{mom2} \\
p^{ab} &=&\frac{\partial \mathcal{L}_{tot}}{\partial B_{ab,0}}=0
\label{mom3} \\
\pi ^{ab} &=&\frac{\partial \mathcal{L}_{tot}}{\partial \gamma _{ab,0}}%
=-\gamma ^{\frac{1}{2}}\left( K^{ab}-\gamma ^{ab}K\right)  \label{mom4} \\
\pi ^{abc} &=&\frac{\partial L_{tot}}{\partial A_{abc,0}}=\frac{1}{6}\gamma
^{\frac{1}{2}}F_{\perp }^{\ \ abc}-\frac{8}{12^{4}}\eta
^{abcd_{1}...d_{7}}A_{d_{1}d_{2}d_{3}}F_{d_{4}...d_{7}}.  \label{mom5}
\end{eqnarray}%
Expressions (\ref{mom1}), (\ref{mom2}) and (\ref{mom3}) are known as \emph{%
primary constraints }\cite{HenneauxTeitelboim}. This means that the
corresponding \textquotedblleft velocities\textquotedblright\ cannot be
expressed in terms of the momenta, and are thus arbitrary.

Now that we have the canonical momenta, we can work out the Hamiltonian for
this system. The canonical Hamiltonian is given by 
\end{subequations}
\begin{equation*}
H_{tot}=\int d^{10}x\left( \alpha _{,0}\pi +\beta _{a,0}\pi ^{a}+\gamma
_{ab,0}\pi ^{ab}+B_{ab,0}p^{ab}+A_{abc,0}\pi ^{abc}-\mathcal{L}_{grav}-%
\mathcal{L}_{form}\right) .
\end{equation*}%
From \cite{BSDeWitt1}, we know that the gravitational Hamiltonian $H_{grav}$
is given by 
\begin{equation}
H_{grav}=\int d^{10}x\left( \alpha _{,0}\pi +\beta _{a,0}\pi ^{a}+\alpha 
\mathcal{H}+\beta _{a}\chi ^{a}\right)  \label{hgrav}
\end{equation}%
with 
\begin{subequations}
\begin{eqnarray}
\mathcal{H} &=&\gamma ^{\frac{1}{2}}\left( K^{ab}K_{ab}-K^{2}-R^{\left(
10\right) }\right) =G_{abcd}\pi ^{ab}\pi ^{cd}-\gamma ^{\frac{1}{2}%
}R^{\left( 10\right) }  \label{hcurlgrav} \\
\chi ^{a} &=&-2\pi _{\ \ |b}^{ab}=-2\pi _{\ \ ,b}^{ab}-\gamma ^{ad}\left(
2\gamma _{bd,c}-\gamma _{bc,d}\right) \pi ^{bc}.  \label{chiigrav}
\end{eqnarray}%
where 
\end{subequations}
\begin{equation}
G_{abcd}=\frac{1}{2}\gamma ^{-\frac{1}{2}}\left( \gamma _{ac}\gamma
_{bd}+\gamma _{ad}\gamma _{bc}-\frac{2}{9}\gamma _{ab}\gamma _{cd}\right)
\label{WDWmetric}
\end{equation}%
is the Wheeler-DeWitt metric.

Consider the remaining part%
\begin{equation}
H_{form}=\int d^{10}x\left( B_{ab,0}p^{ab}+A_{abc,0}\pi ^{abc}-\mathcal{L}%
_{form}\right) .  \label{hform}
\end{equation}%
Due to the constraint (\ref{mom3}), nothing can be done with the first term,
but in $\left( A_{abc,0}\pi ^{abc}-\mathcal{L}_{form}\right) $ we have terms
in $A_{abc,0}$, but these can be expressed in terms of $\pi ^{abc}$ using (%
\ref{mom5}). First, define 
\begin{equation}
\tilde{\pi}^{abc}=\frac{1}{6}\gamma ^{\frac{1}{2}}F_{\perp }^{\ abc}
\label{pitilde1}
\end{equation}%
so that, from (\ref{mom5}), 
\begin{equation}
\tilde{\pi}^{abc}=\pi ^{abc}+\frac{8}{12^{4}}\eta
^{abcd_{1}...d_{7}}A_{d_{1}d_{2}d_{3}}F_{d_{4}...d_{7}}.  \label{pitildeijk}
\end{equation}%
Then from definition of $F_{\perp abc}$ (\ref{fperp1}), we have 
\begin{equation}
A_{bcd,0}=\beta ^{a}F_{abcd}+3\partial _{\lbrack b}B_{cd]}+6\alpha \gamma ^{-%
\frac{1}{2}}\tilde{\pi}_{bcd}.  \label{aijk0}
\end{equation}%
Using (\ref{aijk0}) to substitute $A_{bcd,0}$ for $\pi _{bcd}$, we can write
down the overall Hamiltonian in the form 
\begin{equation}
H_{tot}=\int d^{10}x\left( \alpha _{,0}\pi +\beta _{a,0}\pi
^{a}+B_{ab,0}p^{ab}+\alpha \mathcal{\tilde{H}}+\beta _{a}\tilde{\chi}%
^{a}+B_{ab}\tilde{\chi}^{ab}\right)  \label{hamil1}
\end{equation}%
where 
\begin{subequations}
\begin{eqnarray}
\mathcal{\tilde{H}} &=&\mathcal{H}+\frac{1}{48}\gamma ^{\frac{1}{2}%
}F_{abcd}F^{abcd}+3\gamma ^{-\frac{1}{2}}\tilde{\pi}^{abc}\tilde{\pi}_{abc}
\label{Hcurtilde} \\
\tilde{\chi}^{a} &=&\chi ^{a}+F_{\ bcd}^{a}\tilde{\pi}^{bcd}
\label{chiitilde} \\
\tilde{\chi}^{ab} &=&-3\tilde{\pi}_{\ \ \ ,c}^{abc}+\frac{1}{12^{2}}\eta
^{aba_{3}...a_{10}}\partial _{\lbrack a_{3}}A_{a_{4}a_{5}a_{6]}}\partial
_{\lbrack a_{7}}A_{a_{8}a_{9}a_{10]}}.  \label{chiijtilde}
\end{eqnarray}%
We see from the Hamiltonian (\ref{hamil1}) that the quantities $\alpha ,$ $%
\beta _{a}$ and $B_{ab}$ are arbitrary, so we set the gauge as convenient.

In order for the primary constraints (\ref{mom1})-(\ref{mom5}) to be
consistent with the equations of motion, the time derivatives of $\pi $, $%
\pi ^{a}$ and $p^{ab}$ must vanish. This corresponds to vanishing Poisson
brackets of these momenta with $H_{tot}$. Immediately this leads to the 
\emph{secondary constraints }\cite{HenneauxTeitelboim}\emph{\ } 
\end{subequations}
\begin{subequations}
\begin{eqnarray}
\mathcal{\tilde{H}} &=&0  \label{cons1} \\
\tilde{\chi}^{a} &=&0  \label{cons2} \\
\tilde{\chi}^{ab} &=&0.  \label{cons3}
\end{eqnarray}%
Consequently, the Hamiltonian vanishes on the constraint surface.

The new constraints (\ref{cons1})-(\ref{cons3}) also have to consistent with
the equations of motion. So their Poisson brackets with $H_{tot}$ must
vanish on the constraint surface, or else there will be further constraints.
Calculating Poisson brackets with $H_{tot}$ reduces to working out the
pairwise brackets between the quantities $\mathcal{\tilde{H}}$, $\tilde{\chi}%
^{a}$ and $\tilde{\chi}^{ab}$. The non-vanishing brackets between the
canonical variables are: 
\end{subequations}
\begin{eqnarray*}
\left[ \pi ,\alpha ^{\prime }\right] &=&\delta \left( x,x^{\prime }\right) \
\ \ \ \ \ \ \left[ \pi ^{a},\beta _{b}^{\prime }\right] =\delta _{\
b}^{a}\delta \left( x,x^{\prime }\right) \ \ \ \ \left[ B_{ab},p^{\prime cd}%
\right] =\delta _{\ \ \ \ ab}^{[cd]}\delta \left( x,x^{\prime }\right) \ \ \
\ \  \\
\left[ \gamma _{ab},\pi ^{\prime cd}\right] &=&\delta _{\ \ \ \
ab}^{(cd)}\delta \left( x,x^{\prime }\right) \ \ \ \ \left[ A_{abc},\pi
^{\prime def}\right] =\delta _{\ \ \ \ \ abc}^{\left[ def\right] }\delta
\left( x,x^{\prime }\right)
\end{eqnarray*}%
Here $^{\prime }$ means that a quantity is evaluated at $x^{\prime }$, $%
\delta \left( x,x^{\prime }\right) $ is the $10$-dimensional delta function
and $\delta _{b_{1}...b_{k}}^{a_{1}...a_{k}}=\delta
_{b_{1}}^{a_{1}}...\delta _{b_{k}}^{a_{k}}.$

Before proceeding to the derivation of the Poisson brackets, we note that in
general, the brackets are expressed in terms of generalized functions - $%
\delta $-functions and their derivatives. So the technically correct way to
handle them is to introduce arbitrary test functions and consider the action
of the generalized function on them.

The calculation can be simplified if we notice the following. For an
arbitrary $\Lambda _{ab}$, we have 
\begin{eqnarray}
\delta _{\Lambda }A &=&\left[ A_{def},\int \tilde{\chi}^{\prime ab}\Lambda
_{ab}^{\prime }d^{10}x^{\prime }\right] =-3\int \left[ A_{def},\pi _{\ \ \
,c^{^{\prime }}}^{\prime abc}\right] \Lambda _{ab}^{\prime }d^{10}x^{\prime }
\notag \\
&=&3\Lambda _{\left[ de,f\right] }  \label{chiijapb}
\end{eqnarray}%
So this implies that $\tilde{\chi}^{ab}$ is the generator of the gauge
transformation 
\begin{equation}
\delta _{\Lambda }A=d\Lambda \text{,}  \label{chiijatrans}
\end{equation}%
and hence 
\begin{equation}
\delta _{\Lambda }F=0\text{.}  \label{chiijftrans}
\end{equation}%
Under this transformation, we have 
\begin{eqnarray}
\delta _{\Lambda }\pi ^{def} &=&\left[ \pi ^{def},\int \tilde{\chi}^{\prime
ab}\Lambda _{ab}^{\prime }d^{10}x^{\prime }\right]  \notag \\
&=&\left[ \pi ^{def},\int \frac{4}{12^{3}}\eta ^{abc_{3}...c_{10}}\partial
_{\lbrack c_{3}}^{\prime }A_{c_{4}c_{5}c_{6]}}^{\prime }\partial _{\lbrack
c_{7}}^{\prime }A_{c_{8}c_{9}c_{10]}}^{\prime }\Lambda _{ab}^{\prime
}d^{10}x^{\prime }\right]  \notag \\
&=&-\frac{2}{12^{3}}\eta ^{defabcg_{3}...g_{6}}\partial _{a}\Lambda
_{bc}F_{g_{3}...g_{6}}  \label{chiijpipb1}
\end{eqnarray}%
and hence 
\begin{equation}
\delta _{\Lambda }\tilde{\pi}^{abc}=0\text{.}  \label{deltapitijk}
\end{equation}%
Using (\ref{chiijftrans}) and (\ref{deltapitijk}), it immediately follows
that all brackets involving $\tilde{\chi}^{ab}$ vanish identically, since
relevant terms in each constraint involve only \ $F$ and $\tilde{\pi}^{abc}$.

Consider the brackets with $\tilde{\chi}_{a}$ now. After some index
manipulation it is possible to rewrite $\tilde{\chi}_{a}$ as 
\begin{equation}
\tilde{\chi}_{a}=\chi _{a}+F_{abcd}\pi ^{bcd}-A_{abc}\tilde{\chi}%
^{bc}-3A_{abc}\pi _{\ \ \ ,d}^{bcd}.  \label{chit3}
\end{equation}%
However, $\tilde{\chi}^{bc}$ is also a constraint and moreover all its
brackets with other constraints vanish, so we can replace $\tilde{\chi}_{a}$
by an irreducible constraint $\hat{\chi}_{a}$ given by 
\begin{equation}
\hat{\chi}_{a}=\chi _{a}+F_{abcd}\pi ^{bcd}-3A_{abc}\pi _{\ \ \ ,d}^{bcd}
\label{chihat}
\end{equation}%
It is hence enough to work out the brackets with $\hat{\chi}_{a}$.

In pure gravity, we know from \cite{BSDeWitt1} that $\chi _{a}$ generates
spatial translations. Hence $\hat{\chi}_{a}$ acts on $\gamma _{ab}$ and $\pi
^{ab}$ as a Lie derivative, that is, for arbitrary $\xi ^{a}$ 
\begin{eqnarray*}
\left[ \gamma _{mn},\int \hat{\chi}_{a}^{\prime }\xi ^{\prime
a}d^{10}x^{\prime }\right] &=&\mathcal{L}_{\xi }\gamma _{mn} \\
\left[ \pi ^{mn},\int \hat{\chi}_{a}^{\prime }\xi ^{\prime a}d^{10}x^{\prime
}\right] &=&\mathcal{L}_{\xi }\pi ^{mn}
\end{eqnarray*}%
We can now work out the action of $\hat{\chi}_{a}$ on $A_{mnp}$ and $\pi
^{mnp}$. 
\begin{equation}
\left[ A_{bcd},\int \hat{\chi}_{a}^{\prime }\xi ^{\prime a}d^{10}x^{\prime }%
\right] =\xi ^{a}F_{abcd}+3\partial _{\lbrack b}\left( \xi
^{a}A_{|a|cd]}\right) =\mathcal{L}_{\xi }A_{bcd}  \label{achihatpb}
\end{equation}%
since $A_{bcd}$ is a $3$-form$.$ For $\pi ^{bcd}$ we have 
\begin{eqnarray}
\left[ \pi ^{bcd},\int \hat{\chi}_{a}^{\prime }\xi ^{\prime
a}d^{10}x^{\prime }\right] &=&\partial _{a}\left( \pi ^{bcd}\xi ^{a}\right)
-3\partial _{a}\left( \xi ^{\lbrack b}\pi ^{|a|cd]}\right) +3\xi ^{\lbrack
b}\pi _{\ \ \ \ ,a}^{cd]a}  \notag \\
&=&\mathcal{L}_{\xi }\pi ^{bcd}  \label{pichihatpb}
\end{eqnarray}%
since $\pi ^{bcd}$ is a tensor density of weight $1$. Therefore $-\hat{\chi}%
_{a}$ generates spatial translations, and hence $\hat{\chi}_{a}$ acts as a
Lie derivative. Noting that $\hat{\chi}_{b}$ is a covector and $\mathcal{%
\tilde{H}}$ is a scalar density of weight $1$, we immediately see that 
\begin{eqnarray}
\left[ \hat{\chi}_{b},\int \hat{\chi}_{a}^{\prime }\xi ^{\prime
a}d^{10}x^{\prime }\right] &=&\mathcal{L}_{\xi }\hat{\chi}_{b}=\left( \xi
^{c}\hat{\chi}_{b}\right) _{,c}+\hat{\chi}_{c}\xi _{,b}^{c}
\label{chihatchihatpb} \\
\left[ \mathcal{\tilde{H}},\int \hat{\chi}_{a}^{\prime }\xi ^{\prime
a}d^{10}x^{\prime }\right] &=&\mathcal{L}_{\xi }\mathcal{\tilde{H}}=\left( 
\mathcal{\tilde{H}}\xi ^{c}\right) _{,c}.  \label{htchihatpb}
\end{eqnarray}%
Introducing new test functions $\sigma ^{a}$ and $\sigma $, respectively, we
have 
\begin{eqnarray*}
\int \int \left[ \hat{\chi}_{b},\hat{\chi}_{a}^{\prime }\right] \sigma
^{b}\xi ^{\prime a}d^{10}xd^{10}x^{\prime } &=&\int \hat{\chi}_{c}\left( \xi
_{,b}^{c}\sigma ^{b}-\xi ^{b}\sigma _{,b}^{c}\right) d^{10}x \\
\int \int \left[ \mathcal{\tilde{H}},\hat{\chi}_{a}^{\prime }\right] \sigma
\xi ^{\prime a}d^{10}xd^{10}x^{\prime } &=&-\int \mathcal{\tilde{H}}\xi
^{c}\sigma _{,c}d^{10}x
\end{eqnarray*}%
after integration by parts. This gives that these brackets vanish on the
constraint surface.

We are now only left with the bracket $\left[ \mathcal{\tilde{H}},\mathcal{%
\tilde{H}}^{\prime }\right] $. From \cite{BSDeWitt1}, we already know $\left[
\mathcal{H},\mathcal{H}^{\prime }\right] $, so only need to work out the
brackets $\left[ F^{2},\tilde{\pi}^{2}\right] $ and $\left[ \tilde{\pi}^{2},%
\tilde{\pi}^{2}\right] $, since the other cross-terms vanish. After some
lengthy calculations, which are given in the Appendix, we find that 
\begin{equation}
\left[ \mathcal{\tilde{H}},\mathcal{\tilde{H}}^{\prime }\right] =2\tilde{\chi%
}^{a}\delta _{,a}\left( x,x^{\prime }\right) +\tilde{\chi}_{\ ,a}^{a}\delta
\left( x,x^{\prime }\right)  \label{hchcpb}
\end{equation}%
which is analogous to the untilded expression for $\left[ \mathcal{H},%
\mathcal{H}^{\prime }\right] $ in \cite{BSDeWitt1}. In particular, $\left[ 
\mathcal{\tilde{H}},\mathcal{\tilde{H}}^{\prime }\right] $ vanishes on the
constraint surface.

Hence the full canonical description of the bosonic sector of $11$%
-dimensional supergravity involves only three primary constraints (\ref{mom1}%
)-(\ref{mom3}) and the three corresponding secondary constraints (\ref{cons1}%
)-(\ref{cons3}). These constraints are \emph{first-class constraints} - that
is, their pairwise brackets vanish on the constraint surface, and they
generate gauge transformations \cite{HenneauxTeitelboim}.

Consider now the quantization of this system. Adopting the same view as in 
\cite{BSDeWitt1}, we will take it that any two field operators taken at the
same space-time point commute. This way, the classical consistency
conditions carry over to the quantum case without anomalies. So we can
perform Dirac quantization \cite{HenneauxTeitelboim} of the system. The
constraints then become conditions on the wavefunction $\Psi $: 
\begin{subequations}
\begin{equation}
\mathcal{\tilde{H}}\Psi =0\ \ \hat{\chi}^{a}\Psi =0\ \ \tilde{\chi}^{ab}\Psi
=0  \label{wfcons2}
\end{equation}%
This implies that $H_{tot}\Psi =0$, and hence from the Schr\"{o}dinger
equation, $\partial \Psi /\partial t=0$. The behaviour of the wavefunction $%
\Psi $ is completely determined by these constraints.

\section{Minisuperspace}

In general the wavefunction $\Psi $ is a function on the
infinite-dimensional superspace which consists of $\gamma _{ab}\left(
x\right) $ and $A_{abc}\left( x\right) $ modulo diffeomorphisms and form
gauge transformations. Behaviour in this infinite-dimensional space is
difficult to describe, so it is useful to reduce the number of variables, by
fixing some degrees of freedom. This way the infinite-dimensional superspace
is reduced to a finite-dimensional minisuperspace.

To reduce the number of degrees of freedom in the metric, we consider the
following ansatz for the $11$-dimensional spacetime metric: 
\end{subequations}
\begin{equation}
ds_{11}^{2}=-\alpha \left( t\right) ^{2}dt^{2}+\sum_{i=1}^{n}e^{2X^{i}\left(
t\right) }d\Omega _{i}^{2}.  \label{metric11}
\end{equation}%
Here each $d\Omega _{i}^{2}$ is the metric of a maximally symmetric $a_{i}$%
-dimensional space with radius of curvature $\pm 1$ or $0$. Since the
space-time is $11$-dimensional, we also have a condition $a_{1}+...+a_{n}=10$%
. For each $i$, $e^{X^{i}}$ is the scale factor of each spatial component.
Thus the only remaining degrees of freedom which remain from $\gamma _{ab}$
are the $X^{i}$. Such an ansatz was used before in \cite%
{IvaschukMelnikovZhuk}, among others, to set up a cosmological
minisuperspace model, however here we take these ideas further to write down
exact solutions of this model in certain cases, and we also consider the
case with a non-vanishing $4$-form, which is particularly relevant for
M-theory.

For definiteness, suppose $a_{1}=3$ and consider the following ansatz for
the $4$-form: 
\begin{subequations}
\label{fansatz}
\begin{eqnarray}
F_{\alpha \beta \gamma \delta } &=&\dot{X}^{0}\left( t\right) \hat{%
\varepsilon}_{\alpha \beta \gamma \delta }.  \label{fansatz1} \\
F_{\mu \nu \rho \sigma } &=&0\ \text{otherwise}
\end{eqnarray}%
where $\hat{\varepsilon}_{\alpha \beta \gamma \delta }$ is the volume form
on the $4$-space with metric $ds^{2}=-dt^{2}+d\Omega _{1}^{2}$. A similar
ansatz has been used in \cite{Kaloper:1997yi}. With this ansatz, the degrees
of freedom $A_{abc}$ are reduced to just $X^{0}\left( t\right) $. We use
this notation to explicitly highlight the fact that this degree of freedom
will also be part of our minisuperspace, on par with the gravitational
degrees of freedom $X^{i}$ for $i=1,...,n$.

The second fundamental form $K_{ab}$ is given in this case by 
\end{subequations}
\begin{equation}
K_{ab}=-\frac{1}{2}\alpha ^{-1}\dot{\gamma}_{ab}.  \label{secfun}
\end{equation}%
From the metric ansatz, we immediately get%
\begin{eqnarray*}
K_{ab}K^{ab} &=&\alpha ^{-2}\sum_{i=1}^{n}a_{i}\left( \dot{X}^{i}\right) ^{2}
\\
K^{2} &=&\alpha ^{-2}\dot{V}^{2}.
\end{eqnarray*}%
where we have defined 
\begin{equation}
V=\sum_{i=1}^{n}a_{i}X^{i}.  \label{defv}
\end{equation}
Hence we have 
\begin{equation*}
\gamma ^{\frac{1}{2}}=\hat{\gamma}^{\frac{1}{2}}e^{V}
\end{equation*}%
where $\hat{\gamma}=\det \left( \hat{\gamma}_{ab}\right) $ is the
determinant of the normalized spatial metric $\hat{\gamma}_{ab}$.

With the ansatz (\ref{fansatz1}) for the $4$-form, the Chern-Simons term in
the action (\ref{sugra2}) vanishes and the $F^{2}$ term becomes 
\begin{equation*}
\frac{1}{48}F^{\mu _{1}...\mu _{4}}F_{\mu _{1}...\mu _{4}}=-\frac{1}{2}%
\alpha ^{-2}e^{-2a_{1}X^{1}}\left( \dot{X}^{0}\right) ^{2}.
\end{equation*}

If we assume spatial sections of finite volume, for simplicity we can
normalize this volume to be unity. Thus, rewriting the action in terms of
the new variables $X^{0}$ and $X^{i}$, and integrating out the spatial
integral, we obtain the action for the minisuperspace model $S_{mss}$: 
\begin{equation}
S_{mss}=\int dt\left[ \mu ^{-1}\left( \sum_{i=1}^{n}a_{i}\left( \dot{X}%
^{i}\right) ^{2}-\dot{V}^{2}+\frac{1}{2}e^{-2a_{1}X^{1}}\left( \dot{X}%
^{0}\right) ^{2}\right) +\mu e^{2V}R^{\left( 10\right) }\right]
\label{actionmss}
\end{equation}%
where we have defined 
\begin{equation}
\mu =\alpha e^{-V}.  \label{mjudef}
\end{equation}%
It can be shown explicitly that the equations of motion which are obtained
from this action are equivalent to the equations obtained when our ans\"{a}%
tze for the metric and the $4$-form are substituted into the full field
equations for supergravity. In particular, note that the equation of motion
for $X^{0}$ is 
\begin{equation}
\frac{d}{dt}\left( \alpha ^{-1}e^{V-2a_{1}X^{1}}\dot{X}^{0}\right) =0.
\label{fdoteq}
\end{equation}%
The field equation for the $4$-form is 
\begin{eqnarray*}
\nabla _{\mu }F^{\mu \nu \rho \sigma } &=&\frac{1}{\sqrt{-g}}\partial _{\mu
}\left( \sqrt{-g}F^{\mu \nu \rho \sigma }\right) \\
&=&\alpha ^{-1}e^{-V}\partial _{0}\left( \alpha e^{V}F^{0\nu \rho \sigma
}\right) +\frac{1}{\sqrt{g_{1}}}\partial _{a}\left( g_{1}^{\frac{1}{2}%
}F^{a\nu \rho \sigma }\right) =0
\end{eqnarray*}%
where $g_{1}$ is the determinant of the metric $d\Omega _{1}^{2}$ for the $3$%
-space. The second term in the sum vanishes due to the ansatz (\ref{fansatz1}%
), and the remaining equation is precisely equivalent to (\ref{fdoteq}).

From our metric ansatz (\ref{metric11}), the general form of the spatial
Ricci scalar $R^{\left( 10\right) }$ is 
\begin{equation*}
R^{\left( 10\right) }=\sum_{i=1}^{n}k_{i}a_{i}\left( a_{i}-1\right)
e^{-2X^{i}}
\end{equation*}%
where $k_{i}=\pm 1$ or $0$. With this, the action (\ref{actionmss}) becomes 
\begin{equation}
S_{mss}=\int dt\left[ \mu ^{-1}\left( \sum_{i=1}^{n}a_{i}\left( \dot{X}%
^{i}\right) ^{2}-\dot{V}^{2}+\frac{1}{2}e^{-2a_{1}X^{1}}\left( \dot{X}%
^{0}\right) ^{2}\right) +\sum_{i=1}^{n}\mu k_{i}a_{i}\left( a_{i}-1\right)
e^{2\left( V-X^{i}\right) }\right] .  \label{actionmssr}
\end{equation}%
In the integrand above, we have a quadratic form in the $X^{A}$ for $%
A=0,1,...,n$. Let $G_{AB}$ be the corresponding minisuperspace metric such
that the Lagrangian is given by 
\begin{equation}
L_{mss}=\mu \left( \mu ^{-2}\frac{1}{2}G_{AB}\dot{X}^{A}\dot{X}^{B}-\mathcal{%
V}\right)  \label{langmssg}
\end{equation}%
where $\mathcal{V}$ is the effective potential given by 
\begin{equation*}
\mathcal{V}=-\sum_{i=1}^{n}\mu k_{i}a_{i}\left( a_{i}-1\right) e^{2\left(
V-X^{i}\right) }
\end{equation*}

The Hamiltonian is given by 
\begin{equation}
H=\mu \left( \frac{1}{2}G^{AB}P_{A}P_{B}+\mathcal{V}\right)  \label{hammss}
\end{equation}%
where the $P_{A}$ are conjugate momenta to the $X^{A}$ and $G^{AB}$ is the
inverse metric satisfying $G^{AB}G_{BC}=\delta _{\ C}^{A}$. Hence the
canonical form of the action is%
\begin{equation}
S_{mss}=\int dt\left[ \dot{X}^{A}P_{A}-\mu \left( \frac{1}{2}%
G^{AB}P_{A}P_{B}+\mathcal{V}\right) \right] .  \label{actionmsscan}
\end{equation}%
As in the general case, the Lagrange multiplier $\mu $ enforces the
Hamiltonian constraint $H=0$. First class constraints are generators of
gauge transformations, but in this case the Hamiltonian generates time
reparametrizations, so the gauge transformations in this case are simply
time reparametrizations. Therefore this constraint gives rise to invariance
under time reparametrization \cite{Cavaglia:1996kb}. The gauge
transformations generated by $H$ are given by \cite{Fradkin:1977hw} 
\begin{subequations}
\label{gaugetrans}
\begin{eqnarray*}
\delta X^{i} &=&\varepsilon \left[ X^{i},H\right] \\
\delta P_{i} &=&\varepsilon \left[ P_{i},H\right] \\
\delta \mu &=&\frac{d\varepsilon }{dt}.
\end{eqnarray*}%
We have a freedom of how to choose $\mu $ and there are some natural choices
for $\mu ,$ but for now let us write $\mu $ most generally as 
\end{subequations}
\begin{equation}
\mu =e^{-2f}  \label{gaugefix}
\end{equation}%
for an arbitrary function $f\left( X^{A}\right) $ \cite{IvaschukMelnikovZhuk}%
. The Hamiltonian equations obtained from (\ref{actionmsscan}) in the gauge (%
\ref{gaugefix}) are equivalent to the system with the Hamiltonian $H^{f}$
given by 
\begin{equation}
H^{f}=\frac{1}{2}\left( G^{f}\right) ^{AB}P_{A}P_{B}+e^{-2f}\mathcal{V}
\label{hamf}
\end{equation}%
together with the crucial constraint 
\begin{equation}
H^{f}=0.  \label{hamconst}
\end{equation}%
where 
\begin{equation*}
\left( G^{f}\right) ^{AB}=e^{-2f}G^{AB}.
\end{equation*}%
Thus effectively, different gauge choices corresponds to different conformal
transformations of the minisuperspace metric.

Quantization transforms the Hamiltonian constraint (\ref{hamconst}) into the
Wheeler-DeWitt equation \cite{BSDeWitt1}: 
\begin{equation}
\hat{H}^{f}\Psi ^{f}=0  \label{wdwgauge}
\end{equation}%
where $\hat{H}^{f}$ is the operator corresponding to (\ref{hamf}), and $\Psi
^{f}$ is the wavefunction in the $f$-gauge. A general prescription for the
quantized Hamiltonian operator is 
\begin{equation*}
\hat{H}^{f}=F\left( X\right) \left( -\frac{1}{2}\Delta ^{f}+a\mathcal{R}%
^{f}+e^{-2f}\mathcal{V}\right) 
\end{equation*}%
where $\Delta ^{f}$ is the Laplace-Beltrami operator of the minisuperspace
metric $G^{f},$ $\mathcal{R}^{f}$ is the Ricci scalar of the metric $G^{f}$, 
$a$ is a constant and $F\left( X\right) $ is just some function of the $X^{A}
$. Such an operator but with $F=0$ was first suggested by Hawking and Page 
\cite{Hawking:1985bk}. This operator is covariant under general coordinate
transformations on the minisuperspace, which is the main reason for choosing
this operator ordering when quantizing the Hamiltonian. We want the equation
(\ref{wdwgauge}) to be equivalent for any choice of $f$. This is true if and
only if (\ref{wdwgauge}) is equivalent to 
\begin{equation}
\hat{H}\Psi =0  \label{wdw0gauge}
\end{equation}%
where $\hat{H}=\hat{H}^{f=0}$, $\Psi =\Psi ^{f=0}$. From the theory of
scalar fields in curved spacetimes \cite{Birrell:1982ix} we set 
\begin{equation}
\Psi ^{f}=e^{bf}\Psi .  \label{psigauge}
\end{equation}%
and then in order for (\ref{wdwgauge}) and (\ref{wdw0gauge}) to be
equivalent, it is known that we need $a=\frac{n-1}{8n}$ and $b=\frac{1-n}{2}$%
. To construct a suitable Hilbert space, we need an inner product which
would also be invariant under change of $f$, and also in which $\hat{H}^{f}$
would be hermitian. It turns out that then the measure on the minisuperspace
is given by 
\begin{equation}
d\omega ^{f}=e^{-2f}\sqrt{\left\vert G^{f}\right\vert }dX^{0}...dX^{n}
\label{mssmeasure1}
\end{equation}%
with $\left\vert G^{f}\right\vert =\left\vert \det \left( G_{AB}^{f}\right)
\right\vert $. With this choice of the measure, $F\left( X\right) =e^{2f}$.
Hence overall, 
\begin{equation}
\hat{H}^{f}=e^{2f}\left( -\frac{1}{2}\Delta ^{f}+a\mathcal{R}^{f}+e^{-2f}%
\mathcal{V}\right)   \label{hamopgen}
\end{equation}%
and we can say that $\hat{H}^{f}$ and $\hat{H}$ are related by 
\begin{equation*}
\hat{H}^{f}\Psi =e^{-\frac{n+3}{2}f}\hat{H}\left( e^{\frac{n-1}{2}f}\Psi
\right) .
\end{equation*}%
The (non-gauge fixed) inner product is then~%
\begin{equation}
\left\langle \Psi _{1}^{f},\Psi _{2}^{f}\right\rangle =\int \Psi _{1}^{f\ast
}\Psi _{2}^{f}d\omega ^{f}  \label{mssip}
\end{equation}%
and it is indeed invariant under changes of $f$.

From the inner product (\ref{mssip}) we see that the momentum representation
is of the form 
\begin{equation}
\hat{P}_{A}=-i\left( \partial _{A}+\frac{1}{2\sqrt{G}}\partial _{A}\sqrt{G}+%
\frac{n-1}{2}\partial _{A}f\right)  \label{momrep}
\end{equation}%
where the extra term is chosen such that $\hat{P}_{A}$ is hermitian with
respect to this inner product.

As we have seen, different gauge choices correspond to different definitions
of the time parameter in the system, so in particular there are two gauges
which will be most useful for us:

\begin{itemize}
\item The gauge $f=0$, which corresponds to the choice $\mu =1$ and hence $%
\alpha =e^{V}$. This choice leads to greatly simplified calculations. In 
\cite{CavagliaMoniz}, the time parameter in this gauge is referred to as
gauge proper time, $t_{g}=t-t_{0}$.

\item The gauge $f=\frac{1}{2}V$, which corresponds to the choice $\mu
=e^{-V}$ and hence $\alpha =1$. The time parameter in this gauge is the
cosmic proper time, $t_{c}$ given by 
\begin{equation}
\frac{dt_{c}}{dt}=e^{V}.  \label{ttaureltriv}
\end{equation}
\end{itemize}

\section{Minisuperspace solutions with a trivial $4$-form}

Now consider a special case of the above scenario. Here we will consider
solutions with a trivial $4$-form and we will take the spatial Ricci scalar
to be 
\begin{equation*}
R^{\left( 10\right) }=\frac{1}{2}K^{2}e^{-2X^{1}}
\end{equation*}%
where $K^{2}=2k_{1}a_{1}\left( a_{1}-1\right) $. This means that only one
spatial component of the space has non-vanishing curvature, and the other
components are flat. In particular, this special case encompasses the
scenario where the external $4$-dimensional spacetime has a
Friedmann-Robertson-Walker metric with $k=-1,0,+1$, and the $7$-dimensional
internal space is a Ricci-flat compact manifold. This case is of interest
from a cosmological point of view and also from the point of view of
M-theory special holonomy compactifications. Hence the Lagrangian is now
given as 
\begin{equation*}
L^{f}=\frac{1}{2}G_{ij}^{f}\dot{X}^{i}\dot{X}^{j}+\frac{1}{2}%
K^{2}e^{-2f}e^{2\left( V-X^{1}\right) }
\end{equation*}%
and the corresponding Hamiltonian 
\begin{equation*}
H^{f}=\frac{1}{2}\left( G^{f}\right) ^{ij}P_{i}P_{j}-\frac{1}{2}%
K^{2}e^{-2f}e^{2\left( V-X^{1}\right) }.
\end{equation*}%
for $i,j=1,...,n$. In order to obtain explicit solutions in both classical
and quantum cases, it is necessary to diagonalise the minisuperspace metric $%
G^{f}$ in such a way that $V-X^{1}$ becomes an independent variable. This is
achieved by making the following change of variables: 
\begin{eqnarray*}
Y^{1} &=&b_{1}\left[ \left( a_{1}-1\right) X^{1}+a_{2}X^{2}+...+a_{n}X^{n}%
\right] \\
Y^{2} &=&b_{2}\left[ \left( a_{1}+a_{2}-1\right)
X^{2}+a_{3}X^{3}+...+a_{n}X^{n}\right] \\
Y^{3} &=&b_{3}\left[ \left( a_{1}+a_{2}+a_{3}-1\right)
X^{3}+a_{4}X^{4}+...+a_{n}X^{n}\right] \\
&&... \\
Y^{n} &=&b_{n}\left[ \left( a_{1}+a_{2}+...+a_{n}-1\right) X_{n}\right]
\end{eqnarray*}%
where the coefficients $b_{i}$ are defined by 
\begin{eqnarray*}
b_{1}^{2} &=&2a_{1}\left( a_{1}-1\right) ^{-1} \\
b_{2}^{2} &=&2a_{2}\left( a_{1}-1\right) ^{-1}\left( a_{1}+a_{2}-1\right)
^{-1} \\
&&... \\
b_{n}^{2} &=&2a_{n}\left( a_{1}+...+a_{n-1}-1\right) ^{-1}\left(
a_{1}+...+a_{n}-1\right) ^{-1}.
\end{eqnarray*}%
Then 
\begin{equation*}
G_{ij}^{f}\dot{X}^{i}\dot{X}^{j}=e^{2f}\left[ -\left( \dot{Y}^{1}\right)
^{2}+\left( \dot{Y}^{2}\right) ^{2}+...+\left( \dot{Y}^{n}\right) ^{2}\right]
\end{equation*}%
and moreover 
\begin{equation}
V=\frac{1}{2}\left( b_{1}Y^{1}-b_{2}Y^{2}-...-b_{n}Y^{n}\right) .
\label{Vnewcoordt}
\end{equation}%
We can now write down the Hamiltonian: 
\begin{equation}
H_{mss}=\frac{1}{2}e^{-2f}\left[ \left(
-p_{1}^{2}+p_{2}^{2}+...+p_{n}^{2}\right) -K^{2}e^{2b_{1}^{-1}Y^{1}}\right]
\label{hmsstriv}
\end{equation}%
where the $p_{i}$ are momenta conjugate to the $Y^{i}$. The constraint $%
H_{mss}=0$ becomes simply 
\begin{equation}
\left( -p_{1}^{2}+p_{2}^{2}+...+p_{n}^{2}\right) -K^{2}e^{2b_{1}^{-1}Y^{1}}=0
\label{hamconstriv}
\end{equation}%
Taking into account the constraint (\ref{hamconstriv}), the classical
equations become%
\begin{equation}
\begin{array}{cc}
\dot{p}_{1}=K^{2}b_{1}^{-1}e^{-2f}e^{2b_{1}^{-1}Y^{1}} & \dot{Y}%
^{1}=-e^{-2f}p_{1} \\ 
\dot{p}_{j}=0 & \dot{Y}^{j}=e^{-2f}p_{j}%
\end{array}
\label{eqmotion}
\end{equation}%
where $j=2,...,n$. Since the \textquotedblleft potential\textquotedblright\
does not depend on $Y^{i}$, we get that the $p_{i}$ are constant. Also here
we see that in order to be able to solve these equations easily, it is
convenient to choose a gauge time parameter $\tau $ given by%
\begin{equation}
\tau =\int_{t_{0}}^{t}e^{-2f}dt.  \label{tauparam}
\end{equation}%
Hence in the gauge $f=0$ we have $\tau =t_{g}$. Changing the time parameter,
the equations of motion simplify drastically, becoming 
\begin{equation}
\begin{array}{cc}
\dot{p}_{1}=K^{2}b_{1}^{-1}e^{2b_{1}^{-1}Y^{1}} & \dot{Y}^{1}=-p_{1} \\ 
\dot{p}_{j}=0 & \dot{Y}^{j}=p_{i}%
\end{array}
\label{eqmotionf0}
\end{equation}%
where the dot now denotes time derivatives with respect to $\tau $. These
are the same equations we would get with time parameter $t_{g}$. Thus we get
solutions for $j=2,...,n$:\qquad 
\begin{equation}
Y^{j}=p_{i}\tau +Y_{0}^{j}  \notag
\end{equation}%
where $Y_{0}^{j}$ are constants. Since all momenta except $p_{1}$ are
constant in $\tau $, we can rewrite the Hamiltonian constraint (\ref%
{hamconstriv}) as 
\begin{equation}
K^{2}e^{2b_{1}^{-1}Y^{1}}=\xi ^{2}-p_{1}^{2}  \label{curvmom1}
\end{equation}%
where $\xi $ is a constant given by 
\begin{equation}
\xi ^{2}=p_{2}^{2}+...+p_{n}^{2}.  \label{xisq}
\end{equation}%
Using (\ref{curvmom1}), the equation of motion for $p_{1}$ becomes 
\begin{equation}
\dot{p}_{1}=b_{1}^{-1}\left( \xi ^{2}-p_{1}^{2}\right) .  \label{p1eq1}
\end{equation}

We have thus seen that after a change of variables on the minisuperspace,
the classical minisuperspace system is described by equations (\ref{curvmom1}%
), (\ref{p1eq1}) and the relation between $\dot{Y}^{1}$ and $p_{1}$.
Essentially these are equations of motion of a particle moving in the
potential $-\frac{1}{2}K^{2}e^{2b_{1}^{-1}Y^{1}}$ constrained so that the
total energy vanishes. Apart from the initial conditions, the solutions
depend on the curvature parameter $K^{2}.$ In fact, from (\ref{curvmom1}) we
see that the sign of $K^{2}$ affects the nature of equation (\ref{p1eq1})
and hence the qualitative behaviour of the solution.

A similar system is considered in \cite{Damour:2000hv}-\nocite{Damour:2001sa}%
\cite{Damour:2002et}, where the dynamics of scale factors is studied in the
presence of wall potentials near a cosmological singularity, giving rise to
\textquotedblleft cosmological billiards\textquotedblright .

From (\ref{Vnewcoordt}), the volume factor $e^{V}$ is given by 
\begin{eqnarray}
e^{V} &=&\exp \left[ \frac{1}{2}\left(
b_{1}Y^{1}-b_{2}Y^{2}-...-b_{n}Y^{n}\right) \right]  \notag \\
&=&Ae^{\frac{1}{2}b_{1}Y^{1}}e^{-\frac{1}{2}p_{s}\tau }  \label{evtr}
\end{eqnarray}%
where $A=\exp \left( b_{2}Y_{0}^{2}+...+b_{n}Y_{0}^{n}\right) $ and $%
p_{s}=b_{2}p_{2}+...+b_{n}p_{n}$.

We now proceed to the quantization of the minisuperspace model. The
canonical variables in the minisuperspace are now$\ Y^{i}$ for $i=1,...,n,$
and the corresponding momenta $p_{i}$ for $i=1,...,n$. The minisuperspace
metric is now conformally flat, and is fully flat in the gauge $f=0$.
Eventually we will set $f=0$, so we will disregard terms involving $f$.
Hence in our general expression for the Hamiltonian operator (\ref{hamopgen}%
), the minisuperspace curvature term $\mathcal{R}$ vanishes, and the
Laplace-Beltrami operator reduces to the flat wave operator, and moreover
the expression for momentum operators (\ref{momrep}) reduces to 
\begin{equation*}
\hat{p}_{i}=-i\partial _{Y^{1}}.
\end{equation*}%
Hence the Wheeler-DeWitt equation (\ref{wdwgauge}) for this model is simply%
\begin{equation}
\left( -\partial _{Y^{1}}^{2}+\partial _{Y^{2}}^{2}+...+\partial
_{Y^{n}}^{2}\right) \Psi +K^{2}e^{2b_{1}^{-1}Y^{1}}\Psi =0.
\label{wdweqtrib}
\end{equation}%
This equation separates, and we get 
\begin{equation}
\partial _{Y^{1}}^{2}G_{k_{1}}-\left(
K^{2}e^{2b_{1}^{-1}Y^{1}}-k_{1}^{2}\right) G_{k_{1}}=0  \label{wdweq2}
\end{equation}%
where 
\begin{equation}
\Psi =e^{ik_{2}Y^{2}}...e^{ik_{n}Y^{n}}G_{k_{1}}\left( Y^{1}\right) .
\label{psinongf}
\end{equation}%
Here $k_{1}$ is given by 
\begin{equation}
k_{1}^{2}=k_{2}^{2}+...+k_{n}^{2}  \label{k1sq1}
\end{equation}%
and the $k_{i}$ for $i\geq 2$ are eigenvalues of the momenta $p_{i}$. Note
that (\ref{wdweq2}) is the precise quantum analogue of the classical
constraint (\ref{curvmom1}). Moreover, it can be viewed as a one-dimensional
Schr\"{o}dinger equation with an exponential potential $%
K^{2}e^{2b_{1}^{-1}Y^{1}}$.

Let us now discuss gauge fixing in this system. Consider the following
change of variables: for $j=2,...,n$ let 
\begin{equation}
\xi _{j}=\frac{Y^{j}}{p_{j}}\ \ p_{\xi _{j}}=\frac{1}{2}p_{j}^{2}.
\label{gaugechange}
\end{equation}%
For these variables, the equations of motion become%
\begin{equation*}
\dot{\xi}_{j}=1\ \ \ \dot{p}_{\xi _{j}}=0
\end{equation*}%
and hence the Hamiltonian $H_{mss}$ in these variables is given by%
\begin{equation}
H_{mss}=e^{-2f}\left[ -\frac{1}{2}p_{1}^{2}-\frac{1}{2}%
K^{2}e^{2b_{1}^{-1}Y^{1}}+p_{\xi _{2}}+...+p_{\xi _{n}}\right] .
\label{Hmssnewvar}
\end{equation}%
In the reduced phase space method, we take the gauge choice%
\begin{equation}
\xi _{n}-t=0.  \label{gaugecond}
\end{equation}%
From the equations of motion this further imposes $t=\tau $, and hence $f=0$%
. Hence we get the gauge proper time. The effective Hamiltonian is now%
\begin{equation}
H_{eff}=-\frac{1}{2}p_{1}^{2}-\frac{1}{2}K^{2}e^{2b_{1}^{-1}Y^{1}}+p_{\xi
_{2}}+...+p_{\xi _{n-1}}=-p_{\xi _{n}}  \label{heff1}
\end{equation}%
where the $p_{\xi _{i}}\geq 0$. Thus the gauge-fixed Hamiltonian now does
not vanish in general. So when quantizing we precisely get the Schr\"{o}%
dinger equation%
\begin{equation*}
i\frac{\partial \Psi }{\partial t}=\hat{H}_{eff}\Psi
\end{equation*}%
where 
\begin{equation}
\hat{H}_{eff}=-\left( -\frac{1}{2}\partial _{Y^{1}}^{2}+i\partial _{\xi
_{2}}...+i\partial _{\xi _{n-1}}+\frac{1}{2}K^{2}e^{2b_{1}^{-1}Y^{1}}\right)
\label{heff}
\end{equation}%
and so the solutions are hence%
\begin{equation}
\Psi =e^{ik_{2}^{2}\xi _{2}}...e^{ik_{n-1}^{2}\xi
_{n-1}}e^{ik_{n}^{2}t}G_{k_{1}}\left( Y^{1}\right)  \label{redspacesol}
\end{equation}%
with $G_{k_{1}}\left( Y^{1}\right) $ satisfying (\ref{wdweq2}). So the
gauge-fixed solutions have essentially the same form as (\ref{psinongf}),
but with the condition (\ref{gaugecond}) imposed and with $\xi _{i}=\frac{%
Y_{i}}{k_{i}}$ for $i=2,...,n-1$.

Alternatively, we can use the Faddeev-Popov method. From (\ref{Hmssnewvar}),
the Wheeler-DeWitt equation is 
\begin{equation*}
-\frac{1}{2}\partial _{Y^{1}}^{2}+i\partial _{\xi _{2}}...+i\partial _{\xi
_{n-1}}+i\partial _{\xi _{n}}+\frac{1}{2}K^{2}e^{2b_{1}^{-1}Y^{1}}=0
\end{equation*}%
and the solutions are 
\begin{equation}
\Psi =e^{ik_{2}^{2}\xi _{2}}...e^{ik_{n-1}^{2}\xi _{n-1}}e^{ik_{n}^{2}\xi
_{n}}G_{k_{1}}\left( Y^{1}\right) .  \label{fpsol}
\end{equation}%
The full gauge fixed inner product is given by 
\begin{equation*}
\left\langle \Psi _{1}|\Psi _{2}\right\rangle =\int dY^{1}d\xi _{2}...d\xi
_{n}\Psi _{1}^{\ast }\left( Y^{1},\xi _{2},...,\xi _{n}\right) \delta \left(
\Theta \right) \Delta _{FP}\Psi _{2}\left( Y^{1},\xi _{2},...,\xi _{n}\right)
\end{equation*}%
where $\Theta =0$ is the gauge condition and $\Delta _{FP}$ is the
Faddeev-Popov determinant. For $\Theta =\xi _{n}-t$, which gives the gauge
condition (\ref{gaugecond}), $\Delta _{FP}=1$, so the gauge fixed inner
product is 
\begin{equation}
\left\langle \Psi _{1}|\Psi _{2}\right\rangle =\int dY^{1}d\xi _{2}...d\xi
_{n-1}\Psi _{1}^{\ast }\left( Y^{1},\xi _{2},...,\xi _{n-1},t\right) \Psi
_{2}\left( Y^{1},\xi _{2},...,\xi _{n-1},t\right)  \label{gfip}
\end{equation}%
giving a positive definite Hilbert space. The solutions (\ref{redspacesol})
and (\ref{fpsol}) are basically identical, and the gauge fixed measure
derived using the Faddeev-Popov method is precisely the measure of the
reduced space. Hence the two methods are equivalent.

In any case, the key non-trivial part of a solution of the Wheeler-DeWitt
equation is the function $G_{k_{1}}\left( Y^{1}\right) $ which is a solution
of equation (\ref{wdweq2}), which we can rewrite as an eigenvalue problem 
\begin{equation}
\hat{H}_{Y}G_{k_{1}}=-k_{1}^{2}G_{k_{1}}.  \label{wdweq2ev}
\end{equation}%
where $\hat{H}_{Y}$ is the operator 
\begin{equation*}
\hat{H}_{Y}=\partial _{Y^{1}}^{2}-K^{2}e^{2b_{1}^{-1}Y^{1}}.
\end{equation*}%
Thus operator seems to occur in many different settings, and as such has
been quite well studied. In particular, it appears in Liouville theory \cite%
{Kobayashi:1996kg}, \cite{Fulop:1995di} and also appears in models such as
rolling tachyons \cite{Fredenhagen:2003ut},\cite{Schomerus:2003vv}, \cite%
{Gutperle:2003xf}. Interestingly, various minisuperspace models also contain
the same type of equation, even though the potential is derived from
different perspectives \cite{GasperiniVeneziano96a},\cite%
{GasperiniVeneziano96b},\cite{Cavaglia:1999ka}. Setting $z=\left\vert
K\right\vert b_{1}e^{b_{1}^{-1}Y^{1}}$, we get 
\begin{equation}
z^{2}\frac{\partial ^{2}H}{\partial z^{2}}+z\frac{\partial H}{\partial z}%
-\left( \func{sgn}\left( K^{2}\right) z^{2}-b_{1}^{2}k_{1}^{2}\right) H=0%
\text{.}  \label{wdwbessel}
\end{equation}%
For $K^{2}<0$, this is Bessel's equation with an imaginary parameter, and
for $K^{2}>0$ this is the modified Bessel's equation also with an imaginary
parameter. Hence the solutions of (\ref{wdweq2ev}) are linear combinations
of appropriate Bessel functions.

The operator $\hat{H}_{Y}$ is clearly hermitian on the domain $\mathcal{D}%
_{0}$ of smooth functions with compact support, but this is not enough for a
full definition of a self-adjoint operator. To construct a self-adjoint
operator, we follow the general theory as set out in \cite{ReedSimon2}.
First we take the domain \ $\overline{\mathcal{D}}_{0}$ of the closure of $%
\hat{H}_{Y}$ on $\mathcal{D}_{0}$, and work out the deficiency indices $%
n^{\pm }$ of $\hat{H}_{Y}$. In our case this corresponds to solving equation
(\ref{wdweq2ev}) for eigenvalues $\pm i$ and determining the number of
independent square-integrable solutions in each case. General results say
that there exist self-adjoint extensions of the operator if and only if $%
n^{+}=n^{-}$, and moreover the operator is already self-adjoint if and only
if $n^{\pm }=0$. So to find the self-adjoint extensions we need to solve 
\begin{equation}
\hat{H}_{Y}\phi =\pm i\phi  \label{defeq}
\end{equation}

For $K^{2}>0$, the independent solutions of (\ref{defeq}) are modified
Bessel functions of first and second kind - $I_{\eta ^{\pm }}\left( z\right) 
$ and $K_{\eta ^{\pm }}\left( z\right) $, respectively, with $\eta ^{+}=e^{%
\frac{3}{4}\pi i}$ and $\eta ^{-}=e^{\frac{1}{4}\pi i}$. However all of
these solutions are unbounded, and hence clearly not square-integrable. Thus
in this case, the deficiency indices both vanish, and so the operator $\hat{H%
}_{Y}$ is self-adjoint.

For $K^{2}<0$, the independent solutions of (\ref{defeq}) are 
\begin{equation*}
\phi _{1}^{\pm }\left( z\right) =J_{\eta ^{\pm }}\left( z\right) \ \text{and 
}\phi _{2}^{\pm }\left( z\right) =J_{-\eta ^{\pm }}\left( z\right) \text{,}
\end{equation*}%
where $J_{\nu }\left( z\right) $ are Bessel functions of the first kind. In
this case, only $\phi _{1}^{-}$ and $\phi _{2}^{+}$ are square-integrable,
hence the deficiency indices are $n^{+}=n^{-}=1$, and hence there is a
one-parameter family of self-adjoint extensions defined by 
\begin{equation}
\mathcal{D}_{\theta }=\left\{ \left. \phi +\alpha \left( \phi
_{1}^{-}+e^{2\pi i\theta }\phi _{2}^{+}\right) \right\vert \phi \in 
\overline{\mathcal{D}}_{0}\text{, }\alpha \in \mathbb{C}\right\} 
\label{saextfam}
\end{equation}%
where $\theta \in \left( 0,1\right] $ is the parameter which defines the
extension. As pointed out in \cite{Fredenhagen:2003ut}, the above
prescription for the self-adjoint extension domain defines the asymptotic
behaviour of functions in $\mathcal{D}_{\theta }$ since for $%
Y^{1}\longrightarrow +\infty $, all eigenfunctions of $\hat{H}_{Y}$ have a
slower rate of decay than functions from $\overline{\mathcal{D}}_{0}$.
Therefore the asymptotic behaviour of functions in $\mathcal{D}_{\theta }$
has to be compatible with the asymptotic behaviour of $\phi _{1}^{-}+e^{2\pi
i\theta }\phi _{2}^{+}$ for each $\theta $.

\subsection{Case 1: $K^{2}=0$}

Suppose the spatial curvature fully vanishes, so that $K^{2}=0$. Classically
this gives that $p_{1}$ is also constant, and $Y^{1}$ is given by 
\begin{equation*}
Y^{1}=-p_{1}t+Y_{0}^{1}
\end{equation*}%
for a constant $Y_{0}^{1}$. From (\ref{curvmom1}) we see that $p_{1}^{2}=\xi
^{2}$. Hence in our gauge, the solutions are rather trivial.

Note that from (\ref{ttaureltriv}) that the cosmic and gauge time parameters
are related by 
\begin{equation*}
t_{c}=-\frac{2A}{b_{1}p_{1}+...+b_{n}p_{n}}e^{-\frac{1}{2}%
(b_{1}p_{1}+...+b_{n}p_{n})t}
\end{equation*}%
So depending on the sign of the quantity $\left(
b_{1}p_{1}+...+b_{n}p_{n}\right) $, $t_{c}$ is either always positive for
all values of $t$ or always negative for all values of $t$. Overall, this
can be regarded as a generalization of the Kasner metric. Such solutions
have been obtained many times before - both in a purely gravitational
context \cite{Ryan:1975jw} or as here, a special case of an M-theory model 
\cite{CavagliaMoniz}. To relate to variables $\alpha $,$\beta $ and $\phi $
used in \cite{CavagliaMoniz} and \cite{Cavaglia:2001sb}, set $a_{1}=3$, $%
a_{2}=6$ and $a_{3}=1$, together with 
\begin{eqnarray*}
X^{1} &=&\frac{1}{2}\alpha -\beta -\frac{1}{6}\phi \\
X^{2} &=&-\frac{1}{2}\alpha -\frac{1}{6}\phi \\
X^{3} &=&\alpha +2\beta +\frac{1}{3}\phi .
\end{eqnarray*}%
With these relations, our Lagrange multipliers $\mu $ agree, and hence the
Lagrangian (\ref{langmssg}) becomes 
\begin{equation*}
L_{mss}=\mu ^{-1}\left( 3\dot{\alpha}^{2}-\dot{\phi}^{2}+6\dot{\beta}%
^{2}\right) .
\end{equation*}%
thus precisely as in (\cite{CavagliaMoniz}) and (\cite{Cavaglia:2001sb}).

Back to our variables, in the quantum case, the gauge fixed wavefunctions
which are orthonormal in the gauge fixed measure are 
\begin{equation*}
\Psi _{k_{2},...,k_{n}}\left( Y^{1},\xi _{2},...,\xi _{n-1},t\right) =\left(
2\pi \right) ^{-\frac{n-1}{2}}e^{ik_{1}Y^{1}}e^{ik_{2}^{2}\xi
_{2}}...e^{ik_{n-1}^{2}\xi _{n-1}}e^{ik_{n}^{2}t}
\end{equation*}%
where 
\begin{equation*}
k_{1}^{2}=k_{2}^{2}+...+k_{n}^{2}.
\end{equation*}

\subsection{Case 2: $K^{2}>0$}

Suppose $K^{2}>0.$ From the constraint (\ref{curvmom1}) we see that we must
have $\left\vert p_{1}\right\vert <\xi $. From (\ref{p1eq1}), and using the
condition on $p_{1}$, we get 
\begin{equation*}
b_{1}^{-1}\int dt=\int \frac{dp_{1}}{\xi ^{2}-p_{1}^{2}}=\xi ^{-1}\func{%
arctanh}\left( \xi ^{-1}p_{1}\right)
\end{equation*}%
Hence 
\begin{equation}
p_{1}=\xi \tanh \left( b_{1}^{-1}\xi t+t_{0}\right)  \label{kposp1}
\end{equation}%
Now $Y^{1}$ is determined by 
\begin{equation*}
\dot{Y}^{1}=-\xi \tanh \left( b_{1}^{-1}\xi t+t_{0}\right)
\end{equation*}%
So 
\begin{equation*}
Y^{1}=c_{1}-b_{1}\log \left( \cosh \left( b_{1}^{-1}\xi t+t_{0}\right)
\right)
\end{equation*}%
The relation (\ref{curvmom1}) fixes the constant $c_{1}$, hence $Y^{1}$ is
given by 
\begin{equation}
Y^{1}=-b_{1}\log \left( K\xi ^{-1}\cosh \left( b_{1}^{-1}\xi t+t_{0}\right)
\right)  \label{y1kpossol}
\end{equation}%
Figure \ref{kncnpp} shows the behavior in phase space (with $t_{0}=0$). We
can see that this solution has only one branch - the negative and positive
momentum sectors are smoothly connected.

\FRAME{dtbpFU}{9.1511cm}{6.0934cm}{0pt}{\Qcb{Phase space behaviour for $%
K^{2}>0$ }}{\Qlb{kncnpp}}{fig1}{\special{language "Scientific Word";type
"GRAPHIC";maintain-aspect-ratio TRUE;display "USEDEF";valid_file "F";width
9.1511cm;height 6.0934cm;depth 0pt;original-width 0pt;original-height
0pt;cropleft "0";croptop "1";cropright "1";cropbottom "0";filename
'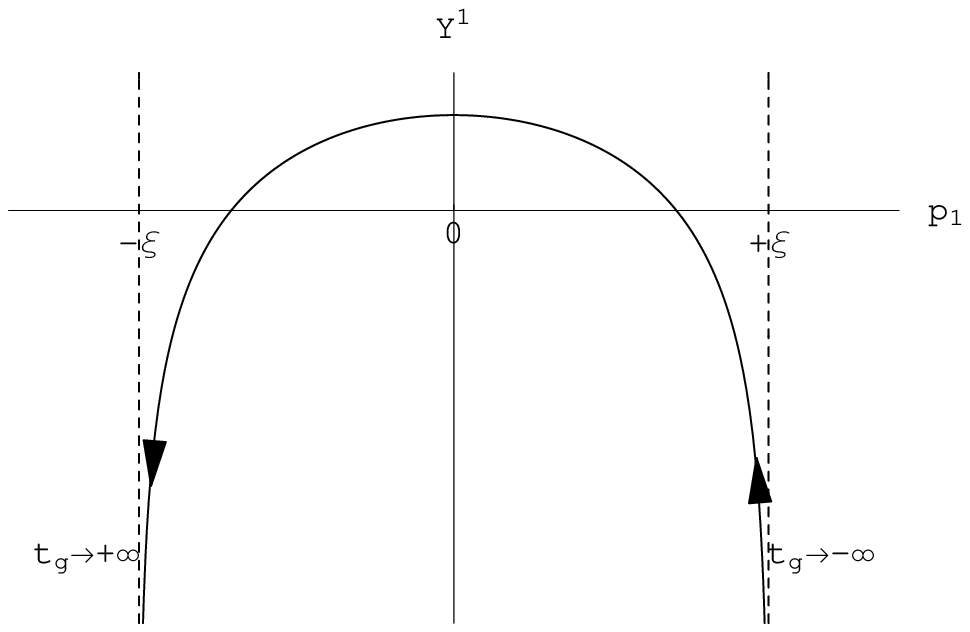';file-properties "XNPEU";}}

We have the following asymptotic behaviour for $Y^{1}$ and $p_{1}$:%
\begin{equation}
\begin{array}{ccc}
t\longrightarrow +\infty & p_{1}\longrightarrow \xi & Y^{1}\sim -\xi
t=-p_{1}t \\ 
t\longrightarrow -\infty & p_{1}\longrightarrow -\xi & Y^{1}\sim +\xi
t=-p_{1}t%
\end{array}
\label{kposasym}
\end{equation}%
Let us now investigate the behaviour of the original scale factors $X^{i}$.
From definition of $Y^{1}$, $X^{1}=V-b_{1}^{-1}Y^{1}$. So let us first look
at the asymptotic behaviour of $V$. From (\ref{Vnewcoordt}), up to a
constant we have 
\begin{equation*}
V=\frac{1}{2}b_{1}Y^{1}-\frac{1}{2}p_{s}t.
\end{equation*}%
Thus from the asymptotic behaviour of $Y_{1}$ (\ref{kposasym}), as $%
t\longrightarrow \pm \infty $ we get 
\begin{equation}
V\sim -\frac{1}{2}t\left( p_{s}\pm b_{1}\xi \right) .  \label{kposvasym}
\end{equation}%
Note that $p_{s}^{2}-b_{1}^{2}\xi ^{2}\leq 0$, but since $b_{1}\xi >0$, we
have $p_{s}-b_{1}\xi <0$ and $p_{s}+b_{1}\xi >0$. Therefore, $%
V\longrightarrow -\infty $ as $t\longrightarrow \pm \infty $, so in fact $V$
has very similar asymptotic behaviour to $Y^{1}$. From (\ref{kposvasym}),
the behaviour of $X^{1}$ is hence easily obtained:%
\begin{eqnarray*}
X^{1} &\sim &-\frac{1}{2}t\left( p_{s}\pm b_{1}\left( 1-2b_{1}^{-2}\right)
\xi \right) \\
&=&-\frac{1}{2}t\left( p_{s}\pm \frac{b_{1}}{a_{1}}\xi \right)
\end{eqnarray*}%
It follows that the qualitative behaviour of $X^{1}$ does actually depend on
the numerical values of the constant momenta $p_{i}$. It can easily be seen
now, that all other $X^{i}$ will also be asymptotically proportional to $t$,
but with different constants of proportionality which also depend on the
initial conditions.

By construction, the overall $11$-dimensional space is Ricci-flat. However
let us look at what happens to the intrinsic curvature from $4$-dimensional
point of view. The expression for the $4$-dimensional Ricci scalar is given
by 
\begin{equation*}
R^{\left( 4\right) }=\frac{1}{2}K^{2}e^{-2X^{1}}+\alpha ^{-2}\left[ \left(
a_{1}-1\right) \overset{\cdot \cdot }{X^{1}}+a_{1}\left( a_{1}-1\right)
\left( \dot{X}^{1}\right) ^{2}\right] .
\end{equation*}%
After changing variables and the time parameter, and applying the constraint
and equations of motion, we get 
\begin{equation}
R^{\left( 4\right) }=e^{-2V}Q\left( p_{1}\right)  \label{ricci4}
\end{equation}%
where $Q\left( p_{1}\right) $ is a quadratic expression in $p_{1}$ with
constant coefficients, the precise form of which is not important here.
Since $p_{1}$ is always bounded (\ref{kposp1}), the curvature blows up when $%
V\longrightarrow -\infty $, and as we know this does happen when $%
t\longrightarrow \pm \infty $. So although the $11$-dimensional space is
flat, from the $4$-dimensional point of view there is a curvature
singularity.

The solutions we had so far were in the gauge time parameter $t$. To relate
it to the cosmic time parameter $t_{c}$, we need to integrate $e^{V}$. In
this case%
\begin{equation*}
e^{V}=A\left[ K^{-1}\xi \func{sech}\left( b_{1}^{-1}\xi t+t_{0}\right) %
\right] ^{\frac{1}{2}b_{1}^{2}}e^{-\frac{1}{2}p_{s}t}
\end{equation*}%
where as before, $p_{s}=b_{2}p_{2}+...+b_{n}p_{n}$. For $t_{0}=0$, the
integral of this expression can be evaluated explicitly in terms of the
hypergeometric function $_{2}F_{1}\left( a,b;c;z\right) $ \cite%
{AbramowitzStegun}: 
\begin{equation*}
t_{c}\left( t\right) =A\left( 2K^{-1}\xi \right) ^{\frac{1}{2}b_{1}^{2}}%
\frac{2}{b_{1}\xi -p_{s}}e^{\frac{1}{2}t\left( b_{1}\xi -p_{s}\right) }\
_{2}F_{1\ }\left( \frac{1}{2}b_{1}^{2},\frac{b_{1}}{4\xi }\left( b_{1}\xi
-p_{s}\right) ;1+\frac{b_{1}}{4\xi }\left( b_{1}\xi -p_{s}\right)
;-e^{2b_{1}^{-1}\xi \tau }\right)
\end{equation*}%
From this we can at least extract asymptotic behaviour of $t_{c}$ as $%
t\longrightarrow \pm \infty $ \cite{AbramowitzStegun}, 
\begin{equation}
t_{c}\sim c_{0}^{\pm }-\frac{2A\left( 2K^{-1}\xi \right) ^{\frac{1}{2}%
b_{1}^{2}}}{p_{s}\pm b_{1}\xi }e^{-\frac{1}{2}t\left( p_{s}\pm b_{1}\xi
\right) }  \label{ttau2}
\end{equation}%
where $c_{0}^{\pm }$ are constants, which we can choose such that $%
c_{0}^{-}=0$. This behaviour is hence similar to the $K^{2}=0$ case for $%
p_{1}=\pm \xi $. We know that $p_{s}-b_{1}\xi <0$ and $p_{s}+b_{1}\xi >0$.
Therefore as $t\longrightarrow -\infty $, the cosmic time parameter $t_{c}$
approaches $0$ from above and as $t\longrightarrow +\infty $, $t_{c}$
approaches $c_{0}^{+}$ from below.

Hence overall, at small $t_{c}$, the overall size of the universe is very
small, and the $4$-dimensional curvature is very high, then as $t_{c}$
increases, the size of the universe increases and hence the curvature
decreases up to a point, after which the universe collapses again and the
curvature blows up within a finite time $c_{0}^{+}$.

Now consider the quantized system. As we already know, for positive $K^{2}$,
the solutions of the Wheeler-DeWitt equation (\ref{wdwbessel}) are modified
Bessel's functions with imaginary parameter $ib_{1}k_{1}$. So the solutions
are linear combinations of functions $I_{ib_{1}k_{1}}\left( z\right) $ and $%
K_{ib_{1}k_{1}}\left( z\right) $. If we impose the condition that the
function be bounded, this uniquely selects $K_{ik}\left( z\right) $ \cite%
{AbramowitzStegun}. This choice selects the the wavefunction which decays as 
$Y^{1}\longrightarrow +\infty $, which is consistent with the exponential
potential in (\ref{wdweq2}). Boundary conditions for this type of
wavefunctions have been well studied \cite{Dabrowski}. From the previous
section, we know that the operator $\hat{H}_{Y}$ is self-adjoint in this
case, and as pointed out in \cite{Fulop:1995di}, this means that there is
only one family of orthogonal eigenfunctions.

The normalized gauge fixed stationary wavefunctions with energy $E=-k_{n}^{2}
$ are 
\begin{equation}
\Psi _{k_{2},...,k_{n}}\left( Y^{1},\xi _{2},...,\xi _{n-1}\right) =\frac{1}{%
\left( 2\pi \right) ^{\frac{n-2}{2}}}\sqrt{\frac{2k_{1}\sinh \pi b_{1}k_{1}}{%
\pi ^{2}}}e^{ik_{2}^{2}\xi _{2}}...e^{ik_{n-1}^{2}\xi
_{n-1}}K_{ib_{1}k_{1}}\left( Kb_{1}e^{b_{1}^{-1}Y^{1}}\right) .
\label{wfkneg}
\end{equation}%
where as before, 
\begin{equation*}
k_{1}^{2}=k_{2}^{2}+...+k_{n}^{2}.
\end{equation*}%
From \cite{Cavaglia:1995bb},\cite{Bateman}, we know that 
\begin{equation}
\int_{0}^{\infty }\frac{dx}{x}K_{i\mu }\left( x\right) K_{i\nu }\left(
x\right) =\frac{\pi ^{2}}{2\mu \sinh \pi \mu }\left[ \delta \left( \mu -\nu
\right) +\delta \left( \mu +\nu \right) \right] .  \label{modbesselip}
\end{equation}%
For definiteness we fix $k_{n}>0\,$and thus the stationary wavefunctions (%
\ref{wfkneg}) are orthonormal in the gauge fixed measure: 
\begin{equation}
\left\langle \Psi _{k_{2},...,k_{n}}|\Psi _{k_{2}^{\prime
},...,k_{n}^{\prime }}\right\rangle =\delta \left( k_{2}-k_{2}^{\prime
}\right) ...\delta \left( k_{n}-k_{n}^{\prime }\right) .  \label{kposnorm}
\end{equation}%
These functions are not in the Hilbert space $\mathcal{H}$ of
square-integrable functions since they show oscillatory behaviour as $%
Y^{1}\longrightarrow -\infty $. This problem is resolved by introducing a
weight distribution $\rho _{i}\left( k_{2},...,k_{n}\right) $, so that the
function%
\begin{equation*}
\Psi _{i}\left( Y^{1},\xi _{2},...,\xi _{n-1}\right) =\int \rho _{i}\left(
k_{2},...,k_{n}\right) \Psi _{k_{2},...,k_{n}}\left( Y^{1},\xi _{2},...,\xi
_{n-1}\right) dk_{2}...dk_{n}
\end{equation*}%
is square integrable. From (\ref{kposnorm}) we see that for each $i$, $\Psi
_{i}$ is indeed in $\mathcal{H}$ if and only if $\rho _{i}\left(
k_{2},...,k_{n}\right) $ is square-integrable in the $k_{2},...,k_{n}$, and
the $\Psi _{i}$ are orthonormal if $\rho _{i}$ are such. If we take $\rho
_{i}$ to be highly peaked around $(k_{2},...,k_{n})$, then the qualitative
behaviour of the corresponding $\Psi _{i}$ can be very well approximated by
the non-smeared function $\Psi _{k_{2},...,k_{n}}$ \cite{MadridRHS}.

Now, a general property of $K_{ib_{1}k_{1}}\left( z\right) $ is that for $%
0<z<\left\vert b_{1}k_{1}\right\vert $, the function is oscillatory, and for 
$z>\left\vert b_{1}k_{1}\right\vert $, the function decays with asymptotic
behaviour as $z\longrightarrow +\infty $ given by 
\begin{equation}
K_{ib_{1}k_{1}}\left( z\right) \sim e^{\left( b_{1}k_{1}-1-i\right) \frac{%
\pi }{2}}\sqrt{\frac{\pi }{2}}z^{-\frac{1}{2}}e^{-z}.  \label{kikasyminf}
\end{equation}%
In our case however, $z=Kb_{1}e^{b_{1}^{-1}Y^{1}}$, so the wavefunction
decays extremely fast for $z>\left\vert b_{1}k_{1}\right\vert $. From the
classical solution (\ref{y1kpossol}), 
\begin{equation*}
z=Kb_{1}e^{b_{1}^{-1}Y^{1}}\leq b_{1}\xi ,
\end{equation*}%
so the region of the minisuperspace where the wavefunction is oscillatory
corresponds to the classically allowed region, and outside it, the
wavefunction amplitude is negligibly small.

For $z\longrightarrow 0$, the asymptotic behaviour of $K_{ib_{1}k_{1}}\left(
z\right) $ is \cite{AbramowitzStegun}: 
\begin{equation}
K_{ib_{1}k_{1}}\left( z\right) \sim \frac{\pi }{2b_{1}k_{1}\sinh \left(
b_{1}k_{1}\pi \right) }\left( \frac{\left( \frac{1}{2}z\right) ^{ib_{1}k_{1}}%
}{\Gamma \left( ib_{1}k_{1}\right) }+\frac{\left( \frac{1}{2}z\right)
^{-ib_{1}k_{1}}}{\Gamma \left( -ib_{1}k_{1}\right) }\right)  \label{kikasym0}
\end{equation}%
Using this, for $Y^{1}\longrightarrow -\infty $, we have the following
asymptotic behaviour for $\Psi $:%
\begin{eqnarray}
\Psi _{k_{2},...,k_{n}} &\sim &Ne^{ik_{2}^{2}\xi _{2}}...e^{ik_{n-1}^{2}\xi
_{n-1}}\left( \frac{\left( \frac{1}{2}Kb_{1}\right) ^{ib_{1}k_{1}}}{\Gamma
\left( ib_{1}k_{1}\right) }e^{ik_{1}Y^{1}}+\frac{\left( \frac{1}{2}%
Kb_{1}\right) ^{-ib_{1}k_{1}}}{\Gamma \left( -ib_{1}k_{1}\right) }%
e^{-ik_{1}Y^{1}}\right)  \label{wfkpos} \\
&=&\Psi ^{\left( -\right) }+\Psi ^{\left( +\right) }  \notag
\end{eqnarray}%
where $N$ is a constant. Thus asymptotically, $\Psi $ splits into
left-moving and right-moving parts, $\Psi ^{\left( -\right) }$ and $\Psi
^{\left( +\right) }$ respectively, with the role of the time-like coordinate
being assigned to $Y^{1}$. These plane waves move along the vector $%
(k_{2},...,k_{n-1})$ in the \textquotedblleft space-like\textquotedblright\
part of the minisuperspace. By applying the $Y^{1}$-momentum operator $%
p_{1}=-i\frac{\partial }{\partial Y^{1}}$, we find that the $p_{1}$
eigenvalue for $\Psi ^{\left( -\right) }$ is $k_{1}$ and the eigenvalue for $%
\Psi ^{\left( +\right) }$ is $-k_{1}$. The constant $k_{1}$ corresponds to
the classical quantity $\xi $ and therefore left-movers correspond to the
sector of the classical solution where $p_{1}>0$ and the right movers
correspond to the sector where $p_{1}<0$. Also, from (\ref{wfkpos}), we can
infer that $\left\vert \Psi \right\vert $ fluctuates with amplitude $%
\left\vert \frac{N}{\Gamma \left( ik_{1}\right) }\right\vert $.

The left-moving waves can be interpreted as reflections of the right-moving
waves. Effectively, such a reflection is a transition from the negative
momentum sector to the positive momentum sector. In the classical system,
these sectors are smoothly connected, and in the quantum system, this is
manifested by the fact that the reflection coefficient between the two plane
waves is $R=\left. \left\vert \Psi ^{\left( +\right) }\right\vert
^{2}\right/ \left\vert \Psi ^{\left( -\right) }\right\vert ^{2}=1$. So in
fact the two sectors can be regarded as reflections of each other at $%
Y^{1}\longrightarrow -\infty $.

A similar behaviour was discussed in \cite{GasperiniVeneziano96b}, in the
context of a four-dimensional gravi-dilaton system with a negative,
specially chosen dilaton potential. Here the smooth branch transition arises
naturally from a positive curvature background since the positive curvature
term in our action gives rise to a negative potential in the Hamiltonian (%
\ref{hmsstriv}).

\subsection{Case 3: $K^{2}<0$}

Now suppose $K^{2}<0$. Letting $\tilde{K}^{2}=-K^{2}$ we thus have from (\ref%
{curvmom1}) 
\begin{equation}
\tilde{K}^{2}e^{2b_{1}^{-1}Y_{1}}=p_{1}^{2}-\xi ^{2}  \label{curvmom2}
\end{equation}

Therefore in this case we have $\left\vert p_{1}\right\vert >\xi $, so from (%
\ref{p1eq1}), and using the condition on $p_{1}$, we get 
\begin{equation*}
b_{1}^{-1}\int dt=\int \frac{dp_{1}}{\xi ^{2}-p_{1}^{2}}=\xi ^{-1}\func{%
arccoth}\left( \xi ^{-1}p_{1}\right)
\end{equation*}%
Hence 
\begin{equation}
p_{1}=\xi \coth \left( b_{1}^{-1}\xi t+t_{0}\right)
\end{equation}%
and $Y^{1}$ is given by 
\begin{equation}
Y^{1}=-b_{1}\log \left( \left\vert \tilde{K}\xi ^{-1}\sinh \left(
b_{1}^{-1}\xi t+t_{0}\right) \right\vert \right)
\end{equation}%
Let $t_{0}=0$. Then the phase space behaviour is shown in Figure \ref{kpcnpp}%
. Now we see that there are two branches - one for which $p_{1}>\xi $ and $t$
is positive, and one for which $p_{1}<-\xi $ and $t$ is negative.

\FRAME{ftbpFU}{9.1511cm}{6.0934cm}{0pt}{\Qcb{Phase space behaviour for $%
K^{2}<0$}}{\Qlb{kpcnpp}}{fig2}{\special{language "Scientific Word";type
"GRAPHIC";maintain-aspect-ratio TRUE;display "USEDEF";valid_file "F";width
9.1511cm;height 6.0934cm;depth 0pt;original-width 0pt;original-height
0pt;cropleft "0";croptop "1";cropright "1";cropbottom "0";filename
'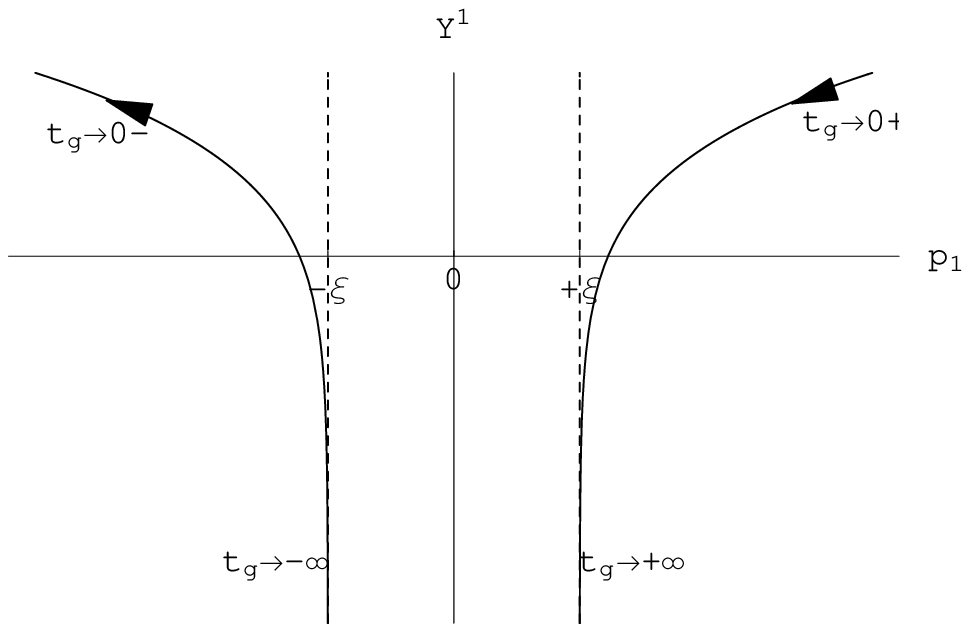';file-properties "XNPEU";}}

The asymptotic behaviour of $Y^{1}$, $V$ and $X^{1}$ as $t\longrightarrow
\pm \infty $ is the same as in the $K^{2}>0$ case, and similarly, $R^{\left(
4\right) }$ blows up when $t\longrightarrow \pm \infty $. To give an
explicit relation between time parameters $t$ and $t_{c}$, we first need 
\begin{equation*}
e^{V}=A\left\vert \tilde{K}^{-1}\xi \func{csch}\left( b_{1}^{-1}\xi
t+t_{0}\right) \right\vert ^{\frac{1}{2}b_{1}^{2}}e^{-\frac{1}{2}p_{s}t}
\end{equation*}%
In this case, at least for $t_{0}=0$, the integral can be explicitly
evaluated in terms of the hypergeometric function $_{2}F_{1}\left(
a,b;c,z\right) $. Hence have \cite{AbramowitzStegun} 
\begin{equation*}
t_{c}\left( t\right) =\left\{ 
\begin{array}{c}
\func{Re}\left( 2A\left( -2\tilde{K}^{-1}\xi \right) ^{\frac{1}{2}b_{1}^{2}}%
\frac{e^{\frac{1}{2}t\left( b_{1}\xi -p_{s}\right) }}{b_{1}\xi -p_{s}}\
_{2}F_{1\ }\left( \frac{1}{2}b_{1}^{2},\frac{b_{1}\left( b_{1}\xi
-p_{s}\right) }{4\xi };1+\frac{b_{1}\left( b_{1}\xi -p_{s}\right) }{4\xi }%
;e^{2b_{1}^{-1}\xi t}\right) \right) \ \text{for }t>0 \\ 
2A\left( 2\tilde{K}^{-1}\xi \right) ^{\frac{1}{2}b_{1}^{2}}\frac{e^{\frac{1}{%
2}t\left( b_{1}\xi -p_{s}\right) }}{b_{1}\xi -p_{s}}\ _{2}F_{1\ }\left( 
\frac{1}{2}b_{1}^{2},\frac{b_{1}\left( b_{1}\xi -p_{s}\right) }{4\xi };1+%
\frac{b_{1}\left( b_{1}\xi -p_{s}\right) }{4\xi };e^{2b_{1}^{-1}\xi
t}\right) \ \text{for }t<0%
\end{array}%
\right.
\end{equation*}%
Using asymptotic behavior of $_{2}F_{1}\left( a,b;b+1,z\right) $ for $%
z\longrightarrow \infty ,$ as $t\longrightarrow \pm \infty $ we get same
behavior as in the positive curvature case \cite{AbramowitzStegun}: 
\begin{equation*}
t_{c}\sim -\frac{2A\left( 2\tilde{K}^{-1}\xi \right) ^{\frac{1}{2}b_{1}^{2}}%
}{p_{s}\pm b_{1}\xi }e^{-\frac{1}{2}t\left( b_{1}p_{s}\pm \xi \right)
}+c_{0}^{\pm }.
\end{equation*}%
where $c_{0}^{\pm }$ are constants, and again we can choose $c_{0}^{-}=0$.
As in the positive curvature case, $p_{s}-b_{1}\xi <0$ and $p_{s}+b_{1}\xi
>0 $. Therefore as $t\longrightarrow -\infty $, $\ t_{c}$ approaches $0$
from above and as $t\longrightarrow +\infty $, $t_{c}$ approaches $c_{0}^{+}$
from below. However we also have this time that as $t\longrightarrow 0$, $%
\left\vert t_{c}\right\vert \longrightarrow \infty $. This implies that $%
c_{0}^{+}$ must actually be negative, and the overall behaviour is that as $%
t $ goes from $-\infty $ to $0$, $t_{c}$ goes from $0$ to $+\infty $, and as 
$t $ goes from $0$ to $+\infty $, $t_{c}$ goes from $-\infty $ to $t_{0}^{+}$%
. Thus, unlike the $K^{2}>0$ case, $t_{c}$ is unbounded.

Hence in this scenario, for negative $t_{c}$ the universe collapses as $%
t_{c}\longrightarrow c_{0}^{+}$ and for positive $t_{c}$ it expands.
Moreover at $t_{c}=0$ and $t_{c}=c_{0}^{+}$, the $4$-dimensional curvature
blows up. Note that the scale factor $X_{n}$ is proportional to $Y_{n}$ and
is hence proportional to $t$. But as $t_{c}\longrightarrow \pm \infty $, $%
t\longrightarrow 0$, thus the volume of this component of the internal space
is stabilised as $t_{c}\longrightarrow \pm \infty $.

Consider the quantum system now. For $K^{2}<0$, the solutions of equation (%
\ref{wdwbessel}) are linear combinations of Bessel functions - $J_{\pm
ib_{1}k_{1}}\left( z\right) $ with an imaginary parameter $\pm ib_{1}k_{1}$,
where $z=\tilde{K}b_{1}e^{b_{1}^{-1}Y^{1}}$. We know that in this case there
is a family of self-adjoint extensions of $\hat{H}_{Y}$ parametrized by $%
\theta \in (0,1]$. Correspondingly there is a set of orthonormal
eigenfunctions $\chi _{k}^{\tilde{\theta}}$ parametrized by some $\tilde{%
\theta}\in (0,1]$ such that there is a one-to-one mapping from $\theta $ to $%
\tilde{\theta}$ \cite{Fredenhagen:2003ut},\cite{Fulop:1995di}. These
eigenfunctions are given by 
\begin{equation}
\chi _{k_{1}}^{\left( \tilde{\theta}\right) }\left( z\right) =\sqrt{\frac{%
k_{1}}{2\sinh \pi b_{1}k_{1}}}\left( J_{-ib_{1}k_{1}}\left( z\right) +\frac{%
\sinh \frac{1}{2}\pi \left( b_{1}k_{1}-2i\tilde{\theta}\right) }{\sinh \frac{%
1}{2}\pi \left( b_{1}k_{1}+2i\tilde{\theta}\right) }J_{ib_{1}k_{1}}\left(
z\right) \right) .  \label{chikpm}
\end{equation}%
Note that for the special case $\tilde{\theta}=\frac{1}{2}$ and $\tilde{%
\theta}=1$, up to normalization, we obtain the functions $%
J_{-ib_{1}k_{1}}\pm J_{ib_{1}k_{1}}$, which when expressed in terms of
Hankel functions, are the orthonormal sets used in \cite{Cavaglia:1996kb}
and \cite{Cavaglia:1999ka}, in particular.

Each of these orthonormal sets is associated with a particular self-adjoint
extension of the Hamiltonian, so once we fix the domain of the operator, we
can only use one particular set of orthonormal eigenfunctions. Again, these
functions are not square-integrable, and as such strictly speaking, do not
belong to $\mathcal{H}$, so as before, to make precise sense of them we need
to smear them with a peaked weight distribution $\rho \,\left(
k_{2},...,k_{n}\right) $, and only then we can say that they belong to a
self-adjoint extension of the Hamiltonian.

So overall, the normalized stationary wavefunctions with energy eigenvalue $%
E=-k_{n}^{2}$ are 
\begin{equation}
\Psi _{k_{2},...,k_{n}}^{\left( \tilde{\theta}\right) }\left( Y^{1},\xi
_{2},...,\xi _{n-1}\right) =\frac{1}{\left( 2\pi \right) ^{\frac{n-2}{2}}}%
e^{ik_{2}^{2}\xi _{2}}...e^{ik_{n-1}^{2}\xi _{n-1}}\chi _{k_{1}}^{\left( 
\tilde{\theta}\right) }\left( Kb_{1}e^{b_{1}^{-1}Y^{1}}\right) .
\end{equation}

As we briefly mentioned before, solutions can also be written in terms of
Hankel functions $H_{ib_{1}k_{1}}^{1}\left( z\right) $ and $%
H_{ib_{1}k_{1}}^{2}\left( z\right) $. Hankel functions are combinations of
Bessel functions and are defined as following: 
\begin{subequations}
\label{hankelbessel}
\begin{eqnarray}
H_{i\nu }^{1}\left( z\right) &=&\frac{e^{\nu \pi }J_{i\nu }\left( z\right)
-J_{-i\nu }\left( z\right) }{\sinh \left( \nu \pi \right) }  \label{h1j} \\
H_{i\nu }^{2}\left( z\right) &=&\frac{J_{-i\nu }\left( z\right) -e^{-\nu \pi
}J_{i\nu }\left( z\right) }{\sinh \left( \nu \pi \right) }.  \label{h2j}
\end{eqnarray}%
Consider the limit as $z\longrightarrow +\infty $ (so that $%
Y^{1}\longrightarrow +\infty $). Then 
\end{subequations}
\begin{eqnarray}
H_{ib_{1}k}^{1}\left( z\right) &\sim &\sqrt{\frac{2}{\pi z}}e^{\frac{%
b_{1}k\pi }{2}}e^{-i\frac{\pi }{4}}e^{iz}  \label{hankel1asyminf} \\
H_{ib_{1}k}^{2}\left( z\right) &\sim &\sqrt{\frac{2}{\pi z}}e^{\frac{%
-b_{1}k\pi }{2}}e^{i\frac{\pi }{4}}e^{-iz}  \label{hankel2asyminf}
\end{eqnarray}%
If we impose the the so-called tunneling boundary condition \cite%
{GasperiniVeneziano96b}, we select only left-moving waves at large $z$ so
that such a solution can be written as 
\begin{equation}
\Psi \left[ k_{2},...,k_{n}\right]
=e^{ik_{2}Y^{2}}...e^{ik_{n}Y^{n}}H_{ib_{1}k_{1}}^{1}\left( \tilde{K}%
b_{1}e^{b_{1}^{-1}Y^{1}}\right) .  \label{wavefunctionkneg}
\end{equation}%
Here the behaviour of the wavefunction is such that for negative $Y^{1},$ $%
\left\vert \Psi \right\vert $ is mostly oscillatory, while for positive $%
Y^{1}$, the wavefunction decays as $e^{-Y^{1}}$, which is much slower than
the decay in the $K^{2}>0$ case. In the current case, all values of $Y^{1}$
are allowed classically, whereas in the $K^{2}>0$ case, $Y^{1}$ is bounded.
This explains the different decay rates.

For $z\longrightarrow 0$, the asymptotic behaviour of $H_{i\nu }^{1}$ is 
\cite{AbramowitzStegun}: 
\begin{equation}
H_{i\nu }^{1}\left( z\right) \sim \frac{1}{\sinh \left( \nu \pi \right) }%
\left( e^{\nu \pi }\frac{\left( \frac{1}{2}z\right) ^{i\nu }}{\Gamma \left(
1+i\nu \right) }-\frac{\left( \frac{1}{2}z\right) ^{-i\nu }}{\Gamma \left(
1-i\nu \right) }\right)  \label{hankelasym0}
\end{equation}%
and therefore, for $v=ib_{1}k_{1}$, we get 
\begin{equation*}
H_{ib_{1}k_{1}}^{1}\left( z\right) \sim \frac{1}{\sinh \left( b_{1}k_{1}\pi
\right) }\left( e^{b_{1}k_{1}\pi }\frac{\left( \frac{1}{2}z\right)
^{ib_{1}k_{1}}}{\Gamma \left( ib_{1}k_{1}\right) }+\frac{\left( \frac{1}{2}%
z\right) ^{-ib_{1}k_{1}}}{\Gamma \left( -ib_{1}k_{1}\right) }\right)
\end{equation*}%
So the asymptotic behaviour as $Y^{1}\longrightarrow -\infty $ is given by 
\begin{eqnarray}
\Psi &=&Ne^{ik_{2}Y^{2}}...e^{ik_{n}Y^{n}}\left( e^{b_{1}k_{1}\pi }\frac{%
\left( \frac{1}{2}\tilde{K}b_{1}\right) ^{ib_{1}k_{1}}}{\Gamma \left(
ib_{1}k_{1}\right) }e^{ik_{1}Y^{1}}+\frac{\left( \frac{1}{2}\tilde{K}%
b_{1}\right) ^{-ib_{1}k_{1}}}{\Gamma \left( -ib_{1}k_{1}\right) }%
e^{-ik_{1}Y^{1}}\right)  \label{wfknega} \\
&=&\Psi ^{\left( -\right) }+\Psi ^{\left( +\right) }  \notag
\end{eqnarray}%
where $N$ is a normalization constant. Interpreting $Y^{1}$ as the timelike
coordinate, the wavefunction is decomposed into left and right moving waves
along the vector $\left( k_{2},...,k_{n}\right) $ in the \textquotedblleft
space-like\textquotedblright\ part of the superspace. Note that $k_{1}$ is
proportional to the magnitude of this vector.

From (\ref{wfknega}), we get that $\left\vert \Psi \right\vert $ oscillates
around $\left\vert \frac{N}{\Gamma \left( ib_{1}k_{1}\right) }\right\vert
e^{b_{1}k_{1}\pi }$ with amplitude $\left\vert \frac{N}{\Gamma \left(
ib_{1}k_{1}\right) }\right\vert $. So although the amplitude of oscillations
is the same as for the case $K^{2}>0$, the fluctuation relative to the value
of $\left\vert \Psi \right\vert $ is very small for large $b_{1}k_{1}$. In
this case the $\Psi ^{\left( -\right) }$ term dominates, and $\left\vert
\Psi \right\vert $ is almost constant as $Y^{1}\longrightarrow -\infty $.

As in the $K^{2}>0$ case the left moving waves correspond to the classical
positive momentum, positive $t$ branch, and can be interpreted as being
incident from the right. The right moving waves correspond to the classical
negative momentum, negative $t$ branch, and can be interpreted as a
reflection of the incident wave. The ratio of the reflected and incident
amplitudes is 
\begin{equation*}
R_{k_{1}}=\frac{\left\vert \Psi ^{\left( +\right) }\right\vert ^{2}}{%
\left\vert \Psi ^{\left( -\right) }\right\vert ^{2}}=e^{-2b_{1}k_{1}\pi }
\end{equation*}%
and this gives the transition probability from the positive momentum branch
to the negative momentum branch. But as we have seen, positive $t$
corresponds to negative $t_{c}$ and vice versa. So we have a transition from
classically disconnected negative time branch to the positive time branch.
Thus there is finite probability of a transition between the two branches
which exhibit very different behaviour. This is similar to results obtained
in \cite{GasperiniVeneziano96b} in a string theory context with a positive
dilaton potential in the Hamiltonian. Here were obtain similar behaviour,
but the potential naturally comes from the spatial curvature. By choosing
the boundary conditions as we did, we made sure that the transition is in
the correct direction when compared with classical solutions.

Although formally we can write down such a solution which exhibits
tunnelling behaviour, it does not mean that it fully makes sense
mathematically. Indeed, if we look at the expression of Hankel functions in
terms of Bessel functions (\ref{hankelbessel}), we see that in order to
construct such a function from orthonormal functions $\chi _{k_{1}}^{\left(
\nu \right) }$, we would need to use $\chi _{k_{1}}^{\left( \nu \right) }$
for at least two different values of $\nu $. However these functions would
lie in different self-adjoint extensions of the Hamiltonian, and thus the
resulting solution (\ref{wavefunctionkneg}) does not lie in any domain where
the Hamiltonian is self-adjoint. Therefore, on the space where (\ref%
{wavefunctionkneg}) belongs, the Hamiltonian is not self-adjoint, and hence
in particular energy is not observable. Moreover, from Stone's theorem \cite%
{ReedSimon1}, we know that quantum dynamics if unitary if and only if the
infinitesimal generator - the Hamiltonian - is self-adjoint. Hence in this
case, with a non-self adjoint Hamiltonian we also lose unitarity of the
system. This is certainly something in need of further investigation,
because it is currently not clear what is the precise physical explanation
for this. 

\section{Minisuperspace solutions for non-trivial $4$-form}

Now let us consider the case with the non-trivial $4$-form. In order to be
able to get solutions explicitly, we fix the spatial Ricci scalar to be 
\begin{equation*}
R^{\left( 10\right) }=\frac{1}{2}K^{2}e^{-2X^{n}}
\end{equation*}%
where $K^{2}=2k_{n}a_{n}\left( a_{n}-1\right) $. This means that the spatial 
$3$-manifold which is part of the external $4$-dimensional spacetime is
necessarily flat. The Lagrangian is now given as 
\begin{equation*}
L^{f}=\frac{1}{2}G_{AB}^{f}\dot{X}^{A}\dot{X}^{B}+\frac{1}{2}%
K^{2}e^{-2f}e^{2\left( V-X^{n}\right) }
\end{equation*}%
and the corresponding Hamiltonian 
\begin{equation*}
H^{f}=\frac{1}{2}\left( G^{f}\right) ^{AB}P_{A}P_{B}-\frac{1}{2}%
K^{2}e^{-2f}e^{2\left( V-X^{n}\right) }.
\end{equation*}%
for $A,B=0,1,...,n$. As before, we need to diagonalise the metric, but
unlike the previous case, here we need $V-X^{n}$ and \ $X^{1}$ to be
independent variables. Notice that in the previous section $Y^{n}$ is
proportional to $X^{n}$. So if in the definitions for $Y^{1}$ and $Y^{n}$ we
replace $X^{1}$ with $X^{n}$ and vice versa, and $a_{1}$ with $a_{n}$, and
vice versa, we get variables which perfectly fit our needs. The other
variables can obviously remain unchanged, but we relabel them for
convenience. Therefore, overall we get the following set of variables: 
\begin{eqnarray*}
Z^{0} &=&X^{0} \\
Z^{1} &=&c_{1}\left( a_{1}+a_{2}+...+a_{n}-1\right) X^{1} \\
Z^{2} &=&c_{2}\left[ a_{1}X^{1}+\left( a_{2}+a_{3}+...+a_{n}-1\right) X^{2}%
\right]  \\
&&... \\
Z^{n-2} &=&c_{n-2}\left[ a_{1}X^{1}+a_{2}X^{2}+...+a_{n-3}X^{n-3}+\left(
a_{n-2}+a_{n-1}+a_{n}-1\right) X^{n-2}\right]  \\
Z^{n-1} &=&c_{n-1}\left[ \left( a_{n}+a_{n-1}-1\right)
X^{n-1}+a_{n-2}X^{n-2}+...+a_{1}X^{1}\right]  \\
Z^{n} &=&c_{n}\left[ \left( a_{n}-1\right)
X^{n}+a_{n-1}X^{n-1}+...+a_{1}X^{1}\right] .
\end{eqnarray*}%
where the coefficients $c_{i}^{2}$ are defined by 
\begin{eqnarray*}
c_{1}^{2} &=&2a_{1}\left( a_{2}+a_{3}+...+a_{n}-1\right) ^{-1}\left(
a_{1}+a_{2}+...+a_{n}-1\right) ^{-1} \\
c_{2}^{2} &=&2a_{2}\left( a_{3}+a_{4}+...+a_{n}-1\right) ^{-1}\left(
a_{2}+a_{3}+...+a_{n}-1\right) ^{-1} \\
&&... \\
c_{n-1}^{2} &=&2a_{n-1}\left( a_{n}-1\right) ^{-1}\left(
a_{n-1}+a_{n}-1\right) ^{-1} \\
c_{n}^{2} &=&2a_{n}\left( a_{n}-1\right) ^{-1}.
\end{eqnarray*}%
With these variables we get 
\begin{equation}
G_{AB}^{f}\dot{X}^{A}\dot{X}^{B}=e^{2f}\left( e^{-\frac{2}{3}%
c_{1}^{-1}Z^{1}}\left( \dot{Z}^{0}\right) ^{2}+\left( \dot{Z}^{1}\right)
^{2}+...+\left( \dot{Z}^{n-1}\right) ^{2}-\left( \dot{Z}^{n}\right)
^{2}\right)   \label{Qformnewcoord}
\end{equation}%
\qquad and moreover 
\begin{equation}
V=-\frac{1}{2}\left( c_{1}Z^{1}+...+c_{n-1}Z^{n-1}-c_{n}Z^{n}\right) 
\label{Vnewcoord}
\end{equation}%
Hence we can write down the Hamiltonian: 
\begin{equation}
H_{mss}=\frac{1}{2}e^{-2f}\left( e^{\frac{2}{3}c_{1}^{-1}Z^{1}}\pi
_{0}^{2}+\pi _{1}^{2}+\pi _{2}^{2}+...+\pi _{n-1}^{2}-\pi
_{n}^{2}-K^{2}e^{2c_{n}^{-1}Z^{n}}\right)   \label{hamil2}
\end{equation}%
where $\pi _{A}$ are the momenta conjugate to $Z^{A}$. With this the
Hamiltonian constraint is 
\begin{equation}
e^{\frac{2}{3}c_{1}^{-1}Z_{1}}\pi _{0}^{2}+\pi _{1}^{2}+\pi _{2}^{2}+...+\pi
_{n-1}^{2}-\pi _{n}^{2}-K^{2}e^{2c_{n}^{-1}Z^{n}}=0.  \label{conscurv1}
\end{equation}%
Taking into account the above constraint (\ref{conscurv1}), the classical
equations of motion are%
\begin{equation}
\begin{array}{cc}
\dot{\pi}_{0}=0 & \dot{Z}^{0}=e^{\frac{2}{3}c_{1}^{-1}Z^{1}} \\ 
\dot{\pi}_{1}=-\frac{1}{3}c_{1}^{-1}e^{\frac{2}{3}c_{1}^{-1}Z^{1}}\pi
_{0}^{2} & \dot{Z}^{1}=\pi _{1} \\ 
\dot{\pi}_{n}=K^{2}c_{n}^{-1}e^{2c_{n}^{-1}Z^{n}} & \dot{Z}^{n}=-\pi _{n} \\ 
\dot{\pi}_{j}=0 & \dot{Z}^{j}=\pi _{i}%
\end{array}
\label{eom1f0}
\end{equation}%
where $j=2,...,n-1$ and the dots denote derivatives with respect to
parameter $\tau $, given by (\ref{tauparam}). For $j=2,...,n$ we immediately
write down\qquad 
\begin{equation}
Z^{j}=\pi _{j}\tau +Z_{0}^{j}  \notag
\end{equation}%
where the $Z_{0}^{j}$ are constants. Similarly as before, we can rewrite the
Hamiltonian constraint (\ref{conscurv1}) as 
\begin{equation}
e^{\frac{2}{3}c_{1}^{-1}Z_{1}}\pi _{0}^{2}+\pi _{1}^{2}+\zeta ^{2}-\pi
_{n}^{2}-K^{2}e^{2c_{n}^{-1}Z^{n}}=0.  \label{curvmom1a}
\end{equation}%
where $\zeta $ is a constant given by 
\begin{equation}
\zeta ^{2}=\pi _{2}^{2}+...+\pi _{n-1}^{2}  \label{xisq1}
\end{equation}%
From the equation of motion, we have 
\begin{eqnarray*}
2\pi _{1}\frac{\partial \pi _{1}}{\partial Z^{1}} &=&-\frac{2}{3}%
c_{1}^{-1}\pi _{0}^{2}e^{\frac{2}{3}c_{1}^{-1}Z^{1}} \\
2\pi _{n}\frac{\partial \pi _{n}}{\partial Z_{n}}
&=&-2K^{2}c_{n}^{-1}e^{2c_{n}^{-1}Z^{n}}
\end{eqnarray*}%
Integrating, we get 
\begin{eqnarray}
\pi _{0}^{2}e^{\frac{2}{3}c_{1}^{-1}Z_{1}} &=&\lambda _{1}^{2}-\pi _{1}^{2}
\label{p1z1eq} \\
K^{2}e^{2c_{n}^{-1}Z_{n}} &=&\lambda _{n}^{2}-\pi _{n}^{2}  \label{pnzneq}
\end{eqnarray}%
substituting these expressions into the constraint, we get 
\begin{equation}
\lambda ^{2}=\lambda _{n}^{2}-\lambda _{1}^{2}.  \label{conschi1chin}
\end{equation}%
Moreover, from equations of motion (\ref{eom1f0}) we have 
\begin{eqnarray}
\dot{\pi}_{1} &=&-\frac{1}{3}c_{1}^{-1}\left( \lambda _{1}^{2}-\pi
_{1}^{2}\right)   \label{p1prim} \\
\dot{\pi}_{n} &=&c_{n}^{-1}\left( \lambda _{n}^{2}-\pi _{n}^{2}\right) .
\label{pnprim}
\end{eqnarray}%
Note that from (\ref{p1z1eq}), we must have $\lambda _{1}^{2}\geq 0$ and
from (\ref{pnzneq}) also has to be non-negative. Moreover, from (\ref{p1z1eq}%
), we can induce that $\left\vert \pi _{1}\right\vert <\lambda _{1}$ and
similarly, $\left\vert \pi _{n}\right\vert <\lambda _{n}$ for $K^{2}>0$, and 
$\left\vert \pi _{n}\right\vert >\lambda _{n}$ for $K^{2}<0$.

Consider what happens to the curvature of the $11$-dimensional spacetime.
From Einstein's equation, we get 
\begin{equation}
R^{\left( 11\right) }=-\frac{1}{6}e^{-2V}\pi _{0}^{2}  \label{curv11}
\end{equation}%
hence the curvature blows up as the volume of the space tends to zero.

We now proceed to the quantization of the minisuperspace model. The
canonical variables in the minisuperspace are now $Z^{A}$ for $A=0,1,...,n,$
and the corresponding momenta $\pi _{A}$ for $A=0,1,...,n$. Unlike the case
with a vanishing $4$-form, even in the gauge $f=0$, the minisuperspace
metric $G_{AB}$ is not flat, as it is given by 
\begin{equation*}
G_{AB}=diag\left( e^{-\frac{2}{3}c_{1}^{-1}Z^{1}},1,...,1,-1\right) 
\end{equation*}%
so in particular in our prescription for the Hamiltonian operator (\ref%
{hamopgen}) both the Laplace-Beltrami operator $\Delta $ and the
minisuperspace Ricci scalar $\mathcal{R}$ are non-trivial. We indeed have 
\begin{eqnarray}
\Delta \Psi  &=&\frac{1}{\sqrt{\left\vert G\right\vert }}\frac{\partial }{%
\partial Z^{A}}\left( \sqrt{\left\vert G\right\vert }G^{AB}\frac{\partial }{%
\partial Z^{B}}\Psi \right)   \notag \\
&=&-\frac{1}{3}c_{1}^{-1}\frac{\partial }{\partial Z^{1}}\Psi +G^{AB}\frac{%
\partial }{\partial Z^{A}}\frac{\partial }{\partial Z^{B}}\Psi 
\label{lapbelt}
\end{eqnarray}%
and since the only non-zero components of $\mathcal{R}_{AB}$ are 
\begin{equation*}
\mathcal{R}_{00}=\frac{1}{9}c^{-2}\ \ \ \ \mathcal{R}_{11}=\frac{1}{9}%
c_{1}^{-2}e^{-\frac{2}{3}c_{1}^{-1}Z^{1}}
\end{equation*}%
we get 
\begin{equation}
\mathcal{R}=G^{AB}\mathcal{R}_{AB}=\frac{2}{9}c_{1}^{-2}.  \label{riccimini}
\end{equation}%
So the Wheeler-DeWitt equation is given by 
\begin{equation}
\left( e^{\frac{2}{3}c_{1}^{-1}Z^{1}}\partial _{Z^{0}}^{2}+\partial
_{Z^{1}}^{2}+...+\partial _{Z^{n-1}}^{2}-\partial _{Z^{n}}^{2}-\frac{1}{3}%
c_{1}^{-1}\partial _{Z^{1}}\right) \Psi +\left( K^{2}e^{2c_{n}^{-1}Z^{n}}-%
\frac{4}{9}ac_{1}^{-2}\right) \Psi =0.  \label{wdweq}
\end{equation}%
Due to the presence the $4$-form term, and the resulting different choice of
operator ordering in the Hamiltonian, this equation differs from equation (%
\ref{wdweqtrib}) in the previous section by new extra terms. The covariant
measure on the minisuperspace is given by 
\begin{equation}
d\omega =dZ^{0}...dZ^{n}e^{-\frac{1}{3}c_{1}^{-1}Z^{1}}  \label{mssmeasure}
\end{equation}%
and hence the momentum operators are now given by 
\begin{eqnarray}
\hat{\pi}_{1} &=&-i\left( \partial _{Z^{1}}-\frac{1}{6}c_{1}^{-1}\right) 
\label{momrep1} \\
\hat{\pi}_{A} &=&-i\partial _{A}\ \ \text{for }A\not=1\text{.}
\label{momrepa}
\end{eqnarray}

The equation (\ref{wdweq}) can be solved by separation of variables, so we
use the following ansatz for $\Psi $: 
\begin{equation}
\Psi =e^{i\kappa _{0}Z^{0}}e^{i\kappa _{2}Z^{2}}...e^{i\kappa
_{n-1}Z^{n-1}}F\left( Z^{1}\right) G\left( Z^{n}\right) .  \label{psinongfn4}
\end{equation}%
With this, the Wheeler-DeWitt equation (\ref{wdweq}) becomes 
\begin{equation*}
-\frac{1}{G}\partial _{Z^{n}}^{2}G+\frac{1}{F}\left( \partial _{Z^{1}}^{2}-%
\frac{1}{3}c_{1}^{-1}\partial _{Z^{1}}\right) F-\kappa _{0}^{2}e^{\frac{2}{3}%
c_{1}^{-1}Z^{1}}+K^{2}e^{2c_{n}^{-1}Z^{n}}=\kappa ^{2}
\end{equation*}%
where 
\begin{equation*}
\kappa ^{2}=\kappa _{2}^{2}+...+\kappa _{n-1}^{2}+\frac{4}{9}ac_{1}^{-2}.
\end{equation*}%
Separating the variables, we get the following equations for $F$ and $G$: 
\begin{eqnarray}
\left( \partial _{Z^{1}}^{2}-\frac{1}{3}c_{1}^{-1}\partial _{Z^{1}}\right)
F-\left( \kappa _{0}^{2}e^{\frac{2}{3}c_{1}^{-1}Z^{1}}-\kappa
_{1}^{2}\right) F &=&0  \label{wdweq2a} \\
\partial _{Z^{n}}^{2}G-\left( K^{2}e^{2c_{n}^{-1}Z^{n}}-\kappa
_{n}^{2}\right) G &=&0  \label{wdweq2b}
\end{eqnarray}%
where $\kappa _{1}$ and $\kappa _{n}$ are constants such that 
\begin{equation}
\kappa _{n}^{2}-\kappa _{1}^{2}=\kappa ^{2}.  \label{quantcons}
\end{equation}%
In order to transform (\ref{wdweq2a}) to the same form as (\ref{wdweq2b}),
set $F=e^{\sigma Z^{1}}\tilde{F}$ for a constant $\sigma $. Turns out that
for $\sigma =\frac{1}{6}c_{1}^{-1}$, we obtain 
\begin{equation}
\partial _{Z^{1}}^{2}\tilde{F}-\left( \kappa _{0}^{2}e^{\frac{2}{3}%
c_{1}^{-1}Z^{1}}-\tilde{\kappa}_{1}^{2}\right) \tilde{F}=0  \label{wdweq2c}
\end{equation}%
where $\tilde{\kappa}_{1}^{2}=\kappa _{1}^{2}-\frac{1}{36}c_{1}^{-2}$. Thus
unsurprisingly, the operator $\hat{H}_{Y}$ appears again. In order to get
bounded solutions of (\ref{wdweq2c}), we need $\tilde{\kappa}_{1}^{2}>0$,
and the solutions are then $K_{3ic_{1}\tilde{\kappa}_{1}}\left( z_{n}\right) 
$ for $z_{n}=Kc_{n}e^{c_{n}^{-1}Z^{n}}$. Now $\kappa _{1}^{2}>0$ and hence
we also have $\kappa _{n}^{2}>0$. Then, as we know, solutions of (\ref%
{wdweq2b}) are Bessel functions $J_{\pm ic_{n}\kappa _{n}}\left(
z_{1}\right) $ or modified Bessel functions $K_{ic_{n}\kappa _{n}}\left(
z_{1}\right) $, for $K^{2}<0$ and $K^{2}>0$ respectively, where $%
z_{1}=3c_{1}\kappa _{0}e^{\frac{1}{3}c_{1}^{-1}Z^{1}}$. As previously, for $%
K^{2}<0$, we get a one-parameter family of self-adjoint extensions (\ref%
{saextfam}) of the Hamiltonian.

Similarly as before, let us discuss gauge fixing. We change variables so
that for $j=2,...,n-1$ we have 
\begin{equation}
\zeta _{j}=\frac{Z^{j}}{\pi _{j}}\ \ \pi _{\zeta _{j}}=\frac{1}{2}\pi
_{j}^{2}
\end{equation}%
and hence the Hamiltonian $H_{mss}$ in these variables is given by%
\begin{equation}
H_{mss}=\frac{1}{2}e^{-2f}\left[ e^{\frac{2}{3}c_{1}^{-1}Z^{1}}\pi
_{0}^{2}+\pi _{1}^{2}-\pi _{n}^{2}-K^{2}e^{2c_{n}^{-1}Z^{n}}\right] +\pi
_{\zeta _{2}}+...+\pi _{\zeta _{n-1}}.
\end{equation}%
In the reduced phase space method, we take the gauge choice $\xi _{n-1}-t=0$%
. From the equations of motion this further imposes $t=\tau $, and hence $f=0
$. Hence we get the gauge proper time. The effective Hamiltonian is now%
\begin{equation}
H_{eff}=\frac{1}{2}\left[ e^{\frac{2}{3}c_{1}^{-1}Z^{1}}\pi _{0}^{2}+\pi
_{1}^{2}-\pi _{n}^{2}-K^{2}e^{2c_{n}^{-1}Z^{n}}\right] +\pi _{\zeta
_{2}}+...+\pi _{\zeta _{n-2}}=-\pi _{\zeta _{n-1}}
\end{equation}%
where the $\pi _{\zeta _{n-1}}\geq 0$. When quantizing, we take the part of
the effective Hamiltonian which depends on $Z^{0},Z^{1}$ and $Z^{n}$ to be
the same as in (\ref{wdweq}), so that overall, we get%
\begin{equation}
\hat{H}_{eff}=-\frac{1}{2}\left( e^{\frac{2}{3}c_{1}^{-1}Z^{1}}\partial
_{Z^{0}}^{2}+\partial _{Z^{1}}^{2}-\partial _{Z^{n}}^{2}\right) +\frac{1}{6}%
c_{1}^{-1}\partial _{Z^{1}}-i\partial _{\zeta _{2}}...-i\partial _{\zeta
_{n-2}}-\frac{1}{2}\left( K^{2}e^{2c_{n}^{-1}Z^{n}}-\frac{4}{9}%
ac_{1}^{-2}\right) 
\end{equation}%
and so the solutions of the corresponding Schr\"{o}dinger equation with $%
E=-\kappa _{n-1}^{2}$ are hence%
\begin{equation}
\Psi =e^{i\kappa _{0}Z^{0}}e^{i\kappa _{2}^{2}\zeta _{2}}...e^{i\kappa
_{n-2}^{2}\zeta _{n-2}}e^{i\kappa _{n-1}^{2}t}e^{\frac{1}{6}c_{1}^{-1}Z^{1}}%
\tilde{F}\left( Z^{1}\right) G\left( Z^{n}\right) 
\end{equation}%
with $\tilde{F}\left( Y^{1}\right) $ satisfying (\ref{wdweq2b}) and $G\left(
Z^{n}\right) $ satisfying (\ref{wdweq2b}). Again, the form of the solution
is same as (\ref{psinongfn4}) but with the gauge condition $\xi _{n-1}-t=0$
imposed and with $\zeta _{i}=\frac{Z^{i}}{\kappa _{i}}$ for $i=0,2,...,n-1$. 

Alternatively, we can use the Faddeev-Popov method. From (\ref{Hmssnewvar}),
the Wheeler-DeWitt equation is 
\begin{equation*}
\hat{H}_{mss}=-\frac{1}{2}\left( e^{\frac{2}{3}c_{1}^{-1}Z^{1}}\partial
_{Z^{0}}^{2}+\partial _{Z^{1}}^{2}-\partial _{Z^{n}}^{2}\right) +\frac{1}{6}%
c_{1}^{-1}\partial _{Z^{1}}-i\partial _{\zeta _{2}}...-i\partial _{\zeta
_{n-2}}-i\partial _{\zeta _{n}-1}-\frac{1}{2}\left(
K^{2}e^{2c_{n}^{-1}Z^{n}}-\frac{4}{9}ac_{1}^{-2}\right)
\end{equation*}%
and the solutions are 
\begin{equation}
\Psi =e^{i\kappa _{0}Z^{0}}e^{i\kappa _{2}^{2}\zeta _{2}}...e^{i\kappa
_{n-1}^{2}\zeta _{n-1}}e^{\frac{1}{6}c_{1}^{-1}Z^{1}}\tilde{F}\left(
Z^{1}\right) G\left( Z^{n}\right) .
\end{equation}%
The full gauge fixed inner product is given by 
\begin{equation*}
\left\langle \Psi _{1}|\Psi _{2}\right\rangle =\int d\omega _{\zeta }\Psi
_{1}^{\ast }\left( Z^{0},Z^{1},Z^{n},\zeta _{2},...,\zeta _{n-1}\right)
\delta \left( \Theta \right) \Delta _{FP}\Psi _{2}\left(
Z^{0},Z^{1},Z^{n},\zeta _{2},...,\zeta _{n-1}\right)
\end{equation*}%
where $\Theta =0$ is the gauge condition, $\Delta _{FP}$ is the
Faddeev-Popov determinant and the measure is 
\begin{equation*}
d\omega _{\zeta }=dZ^{0}dZ^{1}dZ^{n}d\zeta _{2}...d\zeta _{n-2}d\zeta
_{n-1}e^{-\frac{1}{3}c_{1}^{-1}Z^{1}}\text{.}
\end{equation*}
For $\Theta =\zeta _{n}-t$, which gives the gauge condition (\ref{gaugecond}%
), $\Delta _{FP}=1$, so the gauge fixed inner product is 
\begin{equation}
\left\langle \Psi _{1}|\Psi _{2}\right\rangle =\int d\tilde{\omega}_{\zeta
}\Psi _{1}^{\ast }\left( Z^{0},Z^{1},Z^{n},\zeta _{2},...,\zeta
_{n-2},t\right) \Psi _{2}\left( Z^{0},Z^{1},Z^{n},\zeta _{2},...,\zeta
_{n-2},t\right)  \label{nontrivformip}
\end{equation}%
giving a positive definite Hilbert space with the measure 
\begin{equation*}
d\tilde{\omega}_{\zeta }=dZ^{0}dZ^{1}dZ^{n}d\zeta _{2}...d\zeta _{n-2}e^{-%
\frac{1}{3}c_{1}^{-1}Z^{1}.}
\end{equation*}
Again we see that the two methods give equivalent results.

The equations we got here are all very similar to the equations encountered
in the previous section, so we can write down the solutions straight away.
The classical equation for $\pi _{1}$ (\ref{p1prim}) gives 
\begin{equation}
\pi _{1}=-\lambda _{1}\tanh \left( \frac{1}{3}c_{1}^{-1}\lambda
_{1}t+t_{0}\right)   \label{pi1sol}
\end{equation}%
and from the equation of motion for $Z^{1}$ (\ref{eom1}) and the constraint (%
\ref{p1z1eq}), we get the solution for $Z_{1}$:%
\begin{equation}
Z^{1}=-3c_{1}\log \left( \lambda _{1}^{-1}\pi _{0}\cosh \left( \frac{1}{3}%
c_{1\lambda 1}^{-1}t+t_{0}\right) \right) .  \label{z1sol}
\end{equation}%
This is very similar to the solutions for $Y_{1}$ considered in the previous
section for $K^{2}>0$. In particular, in the phase space this solution has a
single branch.

The solutions for $Z_{n}$ are exactly the same as the solutions for $Y_{1}$
in the previous section. Thus for $K^{2}>0$, we have 
\begin{equation}
\pi _{n}=\lambda _{n}\tanh \left( c_{n}^{-1}\lambda _{n}t+t_{1}\right) 
\end{equation}%
and 
\begin{equation}
Z^{n}=-c_{n}\log \left( K\lambda _{n}^{-1}\cosh \left( c_{n}^{-1}\lambda
_{n}t+t_{1}\right) \right) .
\end{equation}%
Similarly, for $K^{2}<0$, we get 
\begin{eqnarray}
\pi _{n} &=&\zeta _{n}\coth \left( c_{n}^{-1}\lambda _{n}t+t_{1}\right)  \\
Z^{n} &=&-c_{n}\log \left( \left\vert \tilde{K}\lambda _{n}^{-1}\sinh \left(
c_{n}^{-1}\lambda _{n}t+t_{1}\right) \right\vert \right) 
\end{eqnarray}%
where $\tilde{K}^{2}=-K^{2}$. In both cases, the asymptotic behaviour as $%
t\longrightarrow \pm \infty $ is 
\begin{eqnarray*}
Z^{1} &\sim &\mp \lambda _{1}t \\
Z^{n} &\sim &\mp \lambda _{n}t.
\end{eqnarray*}%
We know that the volume parameter $V$ is given by 
\begin{equation}
V=\frac{1}{2}\left( c_{n}Z^{n}-c_{1}Z^{1}-\pi _{s}t\right) +const
\label{vol4nt}
\end{equation}%
where 
\begin{equation*}
\pi _{s}=c_{2}\pi _{2}+...+c_{n-1}\pi _{n-1}.
\end{equation*}%
so as $t\longrightarrow \pm \infty $, 
\begin{equation*}
V\sim \pm \frac{1}{2}t\left( c_{1}\lambda _{1}-c_{n}\lambda _{n}\mp \pi
_{s}\right) .
\end{equation*}%
Noting that $X^{1}=\frac{1}{3}Z^{1}$ and $X^{n}=V-c_{n}^{-1}Z^{n}$, we get
the asymptotic behaviour of the original variables $X^{1}$ and $X^{n}$:%
\begin{eqnarray*}
X^{1} &\sim &\mp \frac{1}{3}\lambda _{1}t \\
X^{n} &\sim &\pm \frac{1}{2}t\left( c_{1\lambda 1}-c_{n}\left(
1-2c_{n}^{-1}\right) \lambda _{n}-\pi _{s}\right) .
\end{eqnarray*}%
Thus as $t\longrightarrow \pm \infty $, $X^{1}\longrightarrow -\infty $, and
the qualitative behaviour of $V$ depends on the sign of $c_{\pm
}=c_{1}\lambda _{1}-c_{n}\lambda _{n}\mp \pi _{s}$. Both the $4$-dimensional
and $11$-dimensional curvatures are asymptotically proportional to $e^{-2V}$%
, so the sign of $c_{\pm }$ affects the behaviour of the curvature. Consider
the following example. If $n=2$, then the internal space is $7$-dimensional,
and moreover $\pi _{s}=0$ and $\lambda _{2}=\lambda _{1}$. This immediately
gives us that $c_{\pm }<0$. Hence as $t\longrightarrow \pm \infty $, $%
V\longrightarrow -\infty $ and from (\ref{curv11}) this implies that $R^{11}$
blows up whenever $t\longrightarrow \pm \infty $.

Now look at the solutions of the Wheeler-DeWitt equation for this system (%
\ref{wdweq}). Again, let us look at cases of $K^{2}>0$ and $K^{2}<0$
separately. In the case $K^{2}>0,$ the normalized gauge-fixed stationary
wavefunctions with energy eigenvalue $E=-\kappa _{n-1}^{2}$ are given by 
\begin{equation}
\Psi _{\kappa _{0},\kappa _{1},...,\kappa _{n}}^{pos}=N_{\kappa _{1},\kappa
_{2}}e^{i\kappa _{0}Z^{0}}e^{i\kappa _{2}^{2}\zeta _{2}}...e^{i\kappa
_{n-2}^{2}\zeta _{n-2}}e^{\frac{1}{6}c_{1}^{-1}Z^{1}}K_{3ic_{1}\kappa
_{1}}\left( 3c_{1}\kappa _{0}e^{\frac{1}{3}c_{1}^{-1}Z^{1}}\right)
K_{ic_{n}\kappa _{n}}\left( Kc_{n}e^{c_{n}^{-1}Z^{n}}\right)
\label{wfntrivkp}
\end{equation}%
where 
\begin{equation*}
N_{\kappa _{1},\kappa _{n}}=\frac{2\kappa _{1}\kappa _{n}\sinh 3\pi
c_{1}\kappa _{1}\sinh \pi b_{1}k_{n}}{\left( 2\pi \right) ^{\frac{n-2}{2}%
}\pi ^{2}}.
\end{equation*}%
Similarly as discussed in the case of vanishing $4$-form, these functions
form an orthonormal basis in the inner product (\ref{nontrivformip}).

For $K^{2}<0,$ the normalized wavefunctions are given by 
\begin{equation}
\Psi _{\kappa _{0},\kappa _{1},...,\kappa _{n}}^{neg\ \left( \tilde{\theta}%
\right) }=Ne^{i\kappa _{0}Z^{0}}e^{i\kappa _{2}^{2}\zeta _{2}}...e^{i\kappa
_{n-2}^{2}\zeta _{n-2}}e^{\frac{1}{6}c_{1}^{-1}Z^{1}}K_{3ic_{1}\kappa
_{1}}\left( 3c_{1}\kappa _{0}e^{\frac{1}{3}c_{1}^{-1}Z^{1}}\right) \chi
_{\kappa _{n}}^{\left( \tilde{\theta}\right) }\left(
Kc_{n}e^{c_{n}^{-1}Z^{n}}\right)  \label{wfntrivkn}
\end{equation}%
where $\chi _{\kappa _{n}}^{\left( \tilde{\theta}\right) }$ are the
orthonormal functions given by (\ref{chikpm}). Exactly as in the previous
section, we get a set of normalized wavefunctions for each self-adjoint
extension of the Hamiltonian.

For $K^{2}<0$, if we take the $Z^{n}$ solution to be a Hankel function, then
as in the trivial $4$-form case, for $Z^{n}\longrightarrow -\infty $, we
could decompose the wavefunction into plane waves similarly as in the case
of the trivial $4$-form. Then we would get a non-trivial reflection
probability $R_{\kappa _{n}}=e^{-2c_{n}\kappa _{n}\pi }$ from the
right-moving wave to the left-moving wave, which would correspond to a
transition from the $\pi _{n}>0$ branch to the $\pi _{n}<0$ branch. However,
as in previous situations, such solutions would still belong to a domain
where the Hamiltonian is not self-adjoint.

\section{Concluding remarks}

We have first derived the canonical formulation of the bosonic sector of
eleven dimensional supergravity, together with the complete constraint
algebra. The brackets of the secondary constraints vanish on the constraint
surface, so all constraints are first-class and there are no new tertiary
constraints. When passing to the quantum system, the constraints become
conditions on the wavefunction which govern its behaviour.

By introducing particular ans\"{a}tze for the metric and the $4$-form we
reduced the system to a minisuperspace model with a finite number of degrees
of freedom. In a special case where only one spatial component has
non-vanishing curvature, both the classical and quantum equations can be
solved exactly. In the positive curvature case, whether with or without the $%
4$-form, there is only one branch of the classical solution, where the
universe first expands after starting out from zero size, reaches a maximum
size, and then collapses again within a finite time. When the universe
becomes small, the wavefunction can be written in terms of plane waves
travelling in opposite directions. These waves can be interpreted as being
reflections of one another, but since their coefficients are equal, the
transition probability is $1$. A similar scenario is considered in \cite%
{GasperiniVeneziano96b}, but the effect that there is only one classical
branch of the solution is achieved there by having a negative dilaton
potential in the Hamiltonian, which is hard to motivate in a realistic
superstring theory context.

In the negative curvature case, the classical solutions give two
disconnected branches, one of which is collapsing universe, and the other
branch is an expanding universe. From the $4$-dimensional point of view,
there is a curvature singularity between the two branches. It is possible to
choose boundary conditions such that at $+\infty $ the solution can be
written as a single wave, but at $-\infty $, it splits into two plane waves
going in opposite direction. This yields a non-trivial transition
probability between the branches. However, such a solution does not lie in a
domain where the Hamiltonian is self-adjoint, and hence by choosing such
boundary conditions, we lose the self-adjointness of the Hamiltonian. It
would be interesting to investigate further the physical reasons for this
lost self-adjointness, especially since the operator which appears here is
mathematically the same as in the String Theory minisuperspace models, so
the same problem should arise in those settings as well.

Apart from the self-adjointness problems, we have seen that the curvature
term and the $4$-form term in our minisuperspace models, in terms of
determining the behaviour of the solution, plays the same role as the
dilaton potential in the gravi-dilaton systems derived from string theory.
This is quite remarkable because these terms naturally from the supergravity
action, whereas the dilaton potentials are put in by hand. It would be
interesting to investigate what happens when there is more than one spatial
curvature term. Such an ansatz would be a generalization of the Freund-Rubin
solution of M-theory \cite{FreundRubin}, where the eleven-dimensional space
is of the form $AdS_{4}\times S^{7}$ with particular scale factors for each
component. In particular, if the $3$-space is curved, then there could
possibly be more interaction $4$-form term and the curvature term. Also, in
further work, a less restrictive metric ansatz with a non-trivial moduli
space could be studied, to see how the moduli space parameters evolve and
what is the behaviour of their wavefunctions. In particular, it would be
interesting to study compactifications on manifolds of special holonomy with
time-dependent moduli. This could either involve compactifications on
general $G_{2}$-holonomy manifolds or maybe on a Calabi-Yau space times a
circle. In the latter case, it could be investigated how mirror symmetry 
\cite{Strominger:1996it} is manifested from the point of view of a
minisuperspace quantization.

Study of M-theory minisuperspace models seems to be a promising area where
there is still much left to be uncovered, and which will hopefully aid us in
the quest to further understand M-theory.

\section*{Acknowledgements}

I would like to thank Malcolm Perry for proposing this project and for the
useful discussions and comments throughout its progress, and I would like to
thank Gary Gibbons and the anonymous referees for careful reading of the
manuscript and for making a number of key, insightful remarks. I also
acknowledge funding from EPSRC.

\appendix

\section{Appendix}

In this Appendix we give the details of the calculations involved in
deriving the expression (\ref{hchcpb}) for the Poisson bracket $\left[ 
\mathcal{\tilde{H}},\mathcal{\tilde{H}}\right] $. To do this, we first need
to know the bracket $\left[ \tilde{\pi}^{abc},\tilde{\pi}^{\prime efg}\right]
$. Expanding, we have 
\begin{gather}
\left[ \tilde{\pi}^{abc},\tilde{\pi}^{\prime efg}\right] =\left[ \pi ^{abc}+%
\frac{32}{12^{4}}\eta ^{abcd_{1}...d_{7}}A_{d_{1}d_{2}d_{3}}\partial
_{\lbrack d_{4}}A_{d_{5}d_{6}d_{7}]},\pi ^{\prime efg}+\frac{32}{12^{4}}\eta
^{efga_{1}...a_{7}}A_{a_{1}a_{2}a_{3}}^{\prime }\partial _{\lbrack
a_{4}}^{\prime }A_{a_{5}a_{6}a_{7}]}^{\prime }\right]  \notag \\
=\left[ \pi ^{abc},\frac{32}{12^{4}}\eta
^{efga_{1}...a_{7}}A_{a_{1}a_{2}a_{3}}^{\prime }\partial _{\lbrack
a_{4}}^{\prime }A_{a_{5}a_{6}a_{7}]}^{\prime }\right] +\left[ \frac{32}{%
12^{4}}\eta ^{abcd_{1}...d_{7}}A_{d_{1}d_{2}d_{3}}\partial _{\lbrack
d_{4}}A_{d_{5}d_{6}d_{7}]},\pi ^{\prime efg}\right]  \notag \\
=-\frac{32}{12^{4}}\eta ^{efga_{1}...a_{7}}\left(
A_{a_{1}a_{2}a_{3}}^{\prime }\delta _{\ \ \ a_{5}a_{6}a_{7}}^{abc}\delta
_{,a_{4}^{\prime }}\left( x,x^{\prime }\right) +\delta _{\ \ \
a_{1}a_{2}a_{3}}^{abc}\delta \left( x,x^{\prime }\right) \partial _{\lbrack
a_{4}}^{\prime }A_{a_{5}a_{6}a_{7}]}^{\prime }\right)  \notag \\
+\frac{32}{12^{4}}\eta ^{abcd_{1}...d_{7}}\left( A_{d_{1}d_{2}d_{3}}\delta
_{\ \ \ d_{5}d_{6}d_{7}}^{efg}\delta _{,d_{4}}\left( x,x^{\prime }\right)
+\delta _{\ \ \ d_{1}d_{2}d_{3}}^{efg}\delta \left( x,x^{\prime }\right)
\partial _{\lbrack d_{4}}A_{d_{5}d_{6}d_{7}]}\right)  \notag \\
=\frac{32}{12^{4}}\eta ^{abcefga_{1}...a_{4}}\left(
A_{a_{1}a_{2}a_{3}}^{\prime }\delta _{,a_{4}^{\prime }}\left( x,x^{\prime
}\right) +\delta \left( x,x^{\prime }\right) \partial _{\lbrack
a_{1}}^{\prime }A_{a_{2}a_{3}a_{4}]}^{\prime }\right.  \notag \\
+\left. A_{a_{1}a_{2}a_{3}}\delta _{,a_{4}}\left( x,x^{\prime }\right)
+\delta \left( x,x^{\prime }\right) \partial _{\lbrack
a_{1}}A_{a_{2}a_{3}a_{4}]}\right)  \notag \\
=\frac{32}{12^{4}}\eta ^{abcefga_{1}...a_{4}}\left(
A_{a_{1}a_{2}a_{3}}^{\prime }\delta _{,a_{4}^{\prime }}\left( x,x^{\prime
}\right) +A_{a_{1}a_{2}a_{3}}\delta _{,a_{4}}\left( x,x^{\prime }\right)
+2\delta \left( x,x^{\prime }\right) \partial _{\lbrack
a_{1}}A_{a_{2}a_{3}a_{4}]}\right)  \label{pitpitpb}
\end{gather}

So, for the $\left[ \mathcal{\tilde{H}},\mathcal{\tilde{H}}^{\prime }\right] 
$ bracket, we have:%
\begin{gather}
\left[ \mathcal{\tilde{H}},\mathcal{\tilde{H}}^{\prime }\right] =\left[ 
\mathcal{H}+\frac{1}{48}\gamma ^{\frac{1}{2}}F_{abcd}F^{abcd}+3\gamma ^{-%
\frac{1}{2}}\tilde{\pi}^{abc}\tilde{\pi}_{abc},\mathcal{H}^{\prime }+\frac{1%
}{48}\gamma ^{\prime \frac{1}{2}}F_{abcd}^{\prime }F^{\prime abcd}+3\gamma
^{\prime -\frac{1}{2}}\tilde{\pi}^{\prime abc}\tilde{\pi}_{abc}^{\prime }%
\right]  \notag \\
=\left[ \mathcal{H},\mathcal{H}^{\prime }\right] +\gamma ^{\frac{1}{2}%
}\gamma ^{\prime -\frac{1}{2}}F^{abcd}\tilde{\pi}_{efg}^{\prime }\left[
A_{bcd},\tilde{\pi}^{\prime efg}\right] _{,a}-\gamma ^{\prime \frac{1}{2}%
}\gamma ^{-\frac{1}{2}}F^{\prime abcd}\tilde{\pi}_{efg}\left[
A_{bcd}^{\prime },\tilde{\pi}^{efg}\right] _{,a^{\prime }}  \notag \\
+36\gamma ^{-\frac{1}{2}}\gamma ^{\prime -\frac{1}{2}}\tilde{\pi}_{abc}%
\tilde{\pi}_{efg}^{\prime }\left[ \tilde{\pi}^{abc},\tilde{\pi}^{\prime efg}%
\right]  \notag \\
=\left[ \mathcal{H},\mathcal{H}^{\prime }\right] +\gamma ^{\frac{1}{2}%
}\gamma ^{\prime -\frac{1}{2}}F^{aefg}\tilde{\pi}_{efg}^{\prime }\delta
_{,a}\left( x,x^{\prime }\right) -\gamma ^{\prime \frac{1}{2}}\gamma ^{-%
\frac{1}{2}}F^{\prime aefg}\tilde{\pi}_{efg}\delta _{,a^{\prime }}\left(
x,x^{\prime }\right)  \label{htilhtilpb1} \\
+\frac{8}{12^{2}}\gamma ^{-\frac{1}{2}}\gamma ^{\prime -\frac{1}{2}}\tilde{%
\pi}_{abc}\tilde{\pi}_{efg}^{\prime }\eta
^{abcefga_{1}a_{2}a_{3}a_{4}}\left( A_{a_{1}a_{2}a_{3}}^{\prime }\delta
_{,a_{4}^{\prime }}\left( x,x^{\prime }\right) +A_{a_{1}a_{2}a_{3}}\delta
_{,a_{4}}\left( x,x^{\prime }\right) \right)  \notag
\end{gather}

In the first line cross terms involving $\mathcal{H}$ vanish because the
form terms involve no derivatives of $\gamma _{ab}$ and $\mathcal{H}$ does
not involve any derivatives of $\pi ^{ab}$. Note that the undifferentiated $%
\delta $-function term in $\tilde{\pi}_{abc}\tilde{\pi}_{efg}^{\prime }\left[
\tilde{\pi}^{abc},\tilde{\pi}^{\prime efg}\right] $ vanishes, because $\eta
^{abcefg}\tilde{\pi}_{abc}\tilde{\pi}_{efg}=0$. Let $\xi _{1}$ and $\xi _{2}$
be arbitrary test functions. Then 
\begin{gather}
\int \int \left[ \mathcal{\tilde{H}},\mathcal{\tilde{H}}^{\prime }\right]
\xi _{1}\xi _{2}^{\prime }dxdx^{\prime }=\int \int \left[ \mathcal{H},%
\mathcal{H}^{\prime }\right] \xi _{1}\xi _{2}^{\prime }dxdx^{\prime }
\label{htilhtilintpb1} \\
+\int \int \left( \left( \gamma \gamma ^{\prime -1}\right) ^{\frac{1}{2}%
}F^{a_{1}...a_{4}}\tilde{\pi}_{a_{2}a_{3}a_{4}}^{\prime }\delta
_{,a_{1}}\left( x,x^{\prime }\right) -\left( \gamma ^{\prime }\gamma
^{-1}\right) ^{\frac{1}{2}}F^{\prime a_{1}...a_{4}}\tilde{\pi}%
_{a_{2}a_{3}a_{4}}\delta _{,a_{1}^{\prime }}\left( x,x^{\prime }\right)
\right) \xi _{1}\xi _{2}^{\prime }dxdx^{\prime }  \notag \\
+\frac{8}{12^{2}}\int \int \left( \gamma \gamma ^{\prime }\right) ^{-\frac{1%
}{2}}\tilde{\pi}_{abc}\tilde{\pi}_{efg}^{\prime }\eta
^{abcefga_{1}...a_{4}}\left( A_{a_{1}a_{2}a_{3}}^{\prime }\delta
_{,a_{4}^{\prime }}\left( x,x^{\prime }\right) +A_{a_{1}a_{2}a_{3}}\delta
_{,a_{4}}\left( x,x^{\prime }\right) \right) \xi _{1}\xi _{2}^{\prime
}dxdx^{\prime }  \notag
\end{gather}

We know from \cite{BSDeWitt1} that 
\begin{equation}
\int \int \left[ \mathcal{H},\mathcal{H}^{\prime }\right] \xi _{1}\xi
_{2}^{\prime }dxdx^{\prime }=\int \chi ^{a}\left( \xi _{1}\xi _{2,a}-\xi
_{1,a}\xi _{2}\right) dx  \label{hchcintpb}
\end{equation}

The second line in (\ref{htilhtilintpb1}) becomes%
\begin{eqnarray*}
&&\int \int \left( \gamma ^{\frac{1}{2}}\gamma ^{\prime -\frac{1}{2}}F^{aefg}%
\tilde{\pi}_{efg}^{\prime }\delta _{,a}\left( x,x^{\prime }\right) -\gamma
^{\prime \frac{1}{2}}\gamma ^{-\frac{1}{2}}F^{\prime aefg}\tilde{\pi}%
_{efg}\delta _{,a^{\prime }}\left( x,x^{\prime }\right) \right) \xi _{1}\xi
_{2}^{\prime }dxdx^{\prime } \\
&=&-\int \int \left( \gamma ^{\prime -\frac{1}{2}}\tilde{\pi}_{efg}^{\prime
}\left( \gamma ^{\frac{1}{2}}F^{aefg}\xi _{1}\right) _{,a}\xi _{2}^{\prime
}-\gamma ^{-\frac{1}{2}}\tilde{\pi}_{efg}\left( \gamma ^{\prime \frac{1}{2}%
}F^{\prime aefg}\xi _{2}^{\prime }\right) _{,a^{\prime }}\xi _{1}\right)
\delta \left( x,x^{\prime }\right) dxdx^{\prime } \\
&=&-\int \left( \gamma ^{-\frac{1}{2}}\tilde{\pi}_{efg}\left( \gamma ^{\frac{%
1}{2}}F^{aefg}\xi _{1}\right) _{,a}\xi _{2}-\gamma ^{-\frac{1}{2}}\tilde{\pi}%
_{efg}\left( \gamma ^{\frac{1}{2}}F^{aefg}\xi _{2}\right) _{,a}\xi
_{1}\right) dx \\
&=&\int F_{\ efg}^{a}\tilde{\pi}^{efg}\left( \xi _{1}\xi _{2,a}-\xi
_{1,a}\xi _{2}\right) dx
\end{eqnarray*}%
Now look at the third line in (\ref{htilhtilintpb1}). After integrating by
parts, and integrating out the $\delta $-function we get 
\begin{equation*}
\int \frac{8}{12^{2}}\eta ^{abcefga_{1}a_{2}a_{3}a_{4}}\gamma ^{-1}\tilde{\pi%
}_{abc}\tilde{\pi}_{efg}A_{a_{1}a_{2}a_{3}}\left( \xi _{1}\xi _{2,a_{4}}-\xi
_{1,a_{4}}\xi _{2}\right) dx=0
\end{equation*}%
again because $\eta ^{abcefg}\tilde{\pi}_{abc}\tilde{\pi}_{efg}=0$.

Thus 
\begin{eqnarray}
\int \int \left[ \mathcal{\tilde{H}},\mathcal{\tilde{H}}^{\prime }\right]
\xi _{1}\xi _{2}^{\prime }dxdx^{\prime } &=&\int \left( \chi ^{a}+F_{\
efg}^{a}\tilde{\pi}^{efg}\right) \left( \xi _{1}\xi _{2,a}-\xi _{1,a}\xi
_{2}\right) dx  \notag \\
&=&\int \tilde{\chi}^{a}\left( \xi _{1}\xi _{2,a}-\xi _{1,a}\xi _{2}\right)
dx  \label{htilhtilintpb}
\end{eqnarray}%
Correspondingly, 
\begin{equation*}
\left[ \mathcal{\tilde{H}},\mathcal{\tilde{H}}^{\prime }\right] =2\tilde{\chi%
}^{a}\delta _{,a}\left( x,x^{\prime }\right) +\tilde{\chi}_{\ ,a}^{a}\delta
\left( x,x^{\prime }\right)
\end{equation*}%
which is completely analogous to the untilded expression. In particular, $%
\left[ \mathcal{\tilde{H}},\mathcal{\tilde{H}}^{\prime }\right] $ vanishes
on the constraint surface.

\bibliographystyle{jhep2}
\bibliography{refs2}

\end{document}